\documentclass[twocolumn,secnumarabic,amssymb, nobibnotes, superscriptaddress, nofootinbib,aps,prd]{revtex4-2}
\usepackage[colorlinks,citecolor=blue,linkcolor=blue,anchorcolor=blue,filecolor=blue,
urlcolor=blue]{hyperref}
\usepackage{natbib}

\usepackage{amsmath}
\usepackage{adjustbox}
\usepackage{graphicx}
\usepackage{booktabs}       
\usepackage{appendix}
\usepackage{amsfonts}       
\usepackage{nicefrac}       
\usepackage{microtype}
\usepackage{cleveref}
\usepackage{xcolor}
\usepackage{multirow}
\usepackage{makecell}
\usepackage{array}
\usepackage{orcidlink}
\usepackage{comment}
\usepackage{float}
\usepackage[new]{revdiff}

\begin{document}

\title{Non-Radial Oscillation Modes in Hybrid Stars with Hyperons and Delta Baryons} 
\author{Ishfaq Ahmad Rather~\orcidlink{0000-0001-5930-7179}}
\email{rather@astro.uni-frankfurt.de}
\affiliation{Institut f\"{u}r Theoretische Physik, Goethe Universit\"{a}t,
Max-von-Laue-Str.~1, D-60438 Frankfurt am Main, Germany}
\author{Kau D. Marquez~\orcidlink{0000-0001-5930-7179}} 
\email{Deceased}
\affiliation{Departamento de Física e Laboratório de Computação Científica Avançada e Modelamento (Lab-CCAM), Instituto Tecnológico de Aeronáutica, DCTA, 12228-900 São José dos Campos/SP, Brazil}
\renewcommand{\thefootnote}{\dag} 
\author{Prashant Thakur~\orcidlink{0000-0003-4189-6176}}
\email{p20190072@goa.bits-pilani.ac.in}
\affiliation{Department of Physics, BITS-Pilani, K. K. Birla Goa Campus, 403726 Goa, India}
\author{Odilon Lourenço~\orcidlink{0000-0002-0935-8565}}
\email{odilon.ita@gmail.com}
\affiliation{Departamento de Física e Laboratório de Computação Científica Avançada e Modelamento (Lab-CCAM),
Instituto Tecnológico de Aeronáutica, DCTA, 12228-900 São José dos Campos/SP, Brazil}

\begin{abstract}
 We study the effects of hyperons, delta baryons, and quark matter phase transitions on $f$-mode oscillations in neutron stars. Using the density-dependent relativistic mean-field model (DDME2) for the hadronic phase and the density-dependent quark mass (DDQM) model for the quark phase, we construct hadronic and hybrid equations of state (EoSs) consistent with astrophysical constraints. Including hyperons and delta baryons soften the EoS, reducing maximum mass, while phase transition to the quark matter further softens the EoS, decreasing the speed of sound and hence the maximum mass. We confirm the well-known overestimation of $f$-mode frequencies by the Cowling approximation (by about 10–30\%) compared to full General Relativity calculation, and show that this discrepancy persists across models including hyperons, $\Delta$ baryons, and a phase transition to quark matter. While the discrepancy generally decreases with stellar mass, it increases near the maximum mass in the presence of a phase transition compared to EoSs without this phenomenology. We derive universal relations connecting the frequencies of the $f$-mode to the average density, compactness, and tidal deformability, finding significant deviations due to hyperons and delta baryons. These deviations could provide distinct observational signatures in gravitational wave data, offering new insights into dense matter physics and advancing gravitational wave asteroseismology of neutron star interiors. Empirical relations for mass-scaled and radius-scaled frequencies are also provided, highlighting the importance of GR calculations for accurate modeling. 
 
\end{abstract}
\maketitle

\section{Introduction}

In recent years, our understanding of the universe has expanded as we now observe astronomical events through multiple signals: electromagnetic waves, gravitational waves, neutrinos, and cosmic rays. 
This new era of multimessenger astronomy enables a more comprehensive view of phenomena like neutron star mergers, providing unprecedented insights into the properties of dense matter. 
Neutron star (NS) asteroseismology, in particular, has emerged as a crucial tool for probing the dense matter equation of state (EoS), especially as gravitational wave detections grow in number and precision. 
Landmark events such as GW170817~\cite{PhysRevLett.119.161101, PhysRevLett.121.161101} and GW190425~\cite{LIGOScientific:2020aai} have already provided valuable EoS constraints, while upcoming facilities like LIGO-Virgo-KAGRA, the Einstein Telescope~\cite{2019BAAS...51c.251S, Punturo_2010, 2021arXiv211106990K} and the Cosmic Explorer are set to push these limits even further.

The EoS that governs nuclear matter at the extreme densities attained inside NSs is central to determining their macroscopic structure and properties. 
Though NSs are largely composed of neutrons, a small but crucial fraction of protons, leptons, and possibly other particles is also present in their interiors. 
These degrees of freedom appear to maintain the stability of nuclear matter under chemical equilibrium and charge neutrality conditions, as well as due to energetic considerations, and are very contingent on the dense matter model adopted. 
However, much remains unknown about the EoS and the exact composition of NS interiors due to the complexity of strong interactions, especially at high densities beyond the nuclear saturation density ($n_0$).
At densities surpassing several times $n_0$, exotic particles beyond the usual nucleons (neutrons and protons) are expected to appear. 
Most theoretical models predict that NS matter could comprise the entire spin-1/2 baryon octet, including the hyperons. 
One must observe that the inclusion of hyperons in the EoS, while energetically favorable, has sparked the so-called ``hyperon puzzle'': hyperons soften the EoS, reducing the maximum mass an NS can achieve and potentially conflicting with observations of massive NSs~\cite{2017hspp.confj1002B}.
To address more exotic degrees of freedom, researchers have also considered other particles like kaons and spin-3/2 baryons within relativistic mean-field models. 
Delta ($\Delta$) baryons, for example, are about $30$~\% heavier than nucleons (with a mass of around $1232$~MeV) and are expected to appear at similar densities to hyperons, in the range of 2-3$n_0$~\cite{PhysRevC.67.038801, LI2018234}. 
Studies suggest that with appropriate coupling strengths, delta baryons could indeed make up a significant fraction of NS matter, potentially impacting the EoS and other NS properties~\cite{PhysRevD.102.063008, Xiang:2003qz, Li:2018qaw, Raduta:2021xiz, Malfatti:2020onm, Marquez:2022gmu}.
At even higher densities, a phase transition from hadronic matter to deconfined quark matter may occur, resulting in a hybrid star structure with a quark core surrounded by a hadronic shell. 
This hadron-quark deconfinement transition is a key prediction of quantum chromodynamics (QCD) at extreme densities~\cite{PhysRevLett.119.161101, Capano2020}.

Understanding neutron star oscillation modes has gained increased attention due to recent advancements in multi-messenger astronomy. 
Gravitational wave observatories, such as LIGO and Virgo, have opened new possibilities for detecting the subtle spacetime ripples generated by these oscillations, especially following neutron star mergers. 
These observations complement electromagnetic data from X-ray and radio telescopes, potentially allowing for constraints on neutron star models through observed mode frequencies, damping times, and mode couplings.  
Each type of mode interacts distinctly with the neutron star's dense matter properties, making them sensitive probes of the EoS and phase transitions, such as the potential appearance of hyperons or deconfined quarks in the stellar core. Hence, by analyzing oscillation modes, one can extract information on the internal structure and composition of neutron stars. These oscillation modes effectively act as a spectral fingerprint of the star's static properties, offering a seismological approach to probe otherwise inaccessible regions of dense nuclear matter. Small deviations in static properties, such as the appearance of exotic phases (e.g., hyperons or quark matter) or changes in the crust composition, can lead to measurable shifts in mode frequencies and damping rates, making them highly sensitive to the nature of matter at extreme densities.

When an NS is mechanically perturbed, it exhibits oscillation behaviors that can be classified into radial and non-radial modes. Radial oscillations involve uniform expansion and contraction while maintaining the star’s spherical shape, showing two classes of behavior based on whether they are localized in the dense core or the lower-density outer envelope of the star. These two regions are separated by a ``wall'' in the adiabatic index at the neutron drip point, that is universally tied to the neutron drip density common to all realistic EoS models~\cite{Gondek:1997fd}. 
Although radial modes do not directly emit gravitational waves (GWs), they can interact with non-radial modes, enhancing GW signals~\cite{PhysRevD.73.084010, PhysRevD.75.084038}. For instance, in the post-merger phase of a binary NS collision, a hyper-massive NS may emit GWs at high frequencies (1–4 kHz), which are potentially detectable~\cite{Chirenti_2019}. 

In contrast, non-radial oscillations -- as $f$-modes, associated with fluid oscillations; $g$-modes, driven by compositional gradients; and $p$-modes, which reflect pressure-driven oscillations -- cause distortions due to forces like pressure and buoyancy~\cite{Kokkotas:1999bd}. Gravitational perturbations in spherically symmetric stars are categorized as polar or axial. Polar perturbations lead to the $f$, $p$, and $g$ modes, while axial perturbations result in the $r$ and $w$ modes. In non-rotating stars, these perturbations are entirely independent~\cite{Benhar:2004xg}. Among non-radial modes, the $f$-mode is particularly significant as it emits detectable GWs. Advanced detectors like the Einstein Telescope and Cosmic Explorer, and possibly even current detectors such as LIGO/Virgo/KAGRA, are expected to observe these signals~\cite{2019BAAS...51c.251S, Punturo_2010, 2021arXiv211106990K, PhysRevLett.122.061104, 10.1093/ptep/ptac073}. The $f$-mode frequency is closely tied to tidal deformability during the inspiral phase of NS mergers, as fluid perturbations peak at the stellar surface, strongly coupling to the tidal field. Apart from neutron star mergers various phenomena can trigger the excitation of $f$-modes in neutron stars, including the formation of newly born neutron stars~\cite{Ferrari:2002ut}, starquakes~\cite{Mock_1998, Kokkotas:1999mn}, magnetar activity~\cite{LIGOScientific:2019ccu, LIGOScientific:2022sts}. For GW170817, the 90\% credible interval for the $f$-mode frequency was estimated between 1.43~kHz and 2.90~kHz for the more massive NS and 1.48~kHz and 3.18~kHz for the less massive one~\cite{Kunjipurayil:2022zah}. Additionally, the $f$-mode relates to NS properties like compactness~\cite{Andersson:1997rn}, moment of inertia~\cite{Lau:2009bu}, and static tidal polarizability~\cite{PhysRevD.90.124023}. These relations are universal, applying even to quark stars without crusts or hybrid stars with first-order transitions~\cite{Zhao:2022tcw}.

The study of the $f$-mode is often conducted using the Cowling approximation instead of a full General Relativistic~(GR) framework. 
In the Cowling approximation, gravitational potential perturbations are neglected, focusing solely on fluid perturbations. This simplification aids calculations but introduces an error of about 10-30\% in the $f$-mode frequency~\cite{Pradhan:2020amo, Kunjipurayil:2022zah}. On the other hand, the full GR framework incorporates both fluid and metric perturbations, comprehensively addressing the limitations of the Cowling approximation.
The Cowling approximation neglects metric perturbations, leading to smaller errors for neutron stars with higher masses. This is because massive neutron stars have fluid perturbations that peak more strongly near the surface, while their weaker core coupling to metric perturbations reduces the impact of these neglected terms. Consequently, the relative error between the Cowling approximation and the full GR framework decreases as the mass of the neutron star increases. 
Several studies have explored the $f$-mode oscillations of neutron stars under the Cowling approximation, considering nucleonic and hyperonic compositions~\cite{Pradhan:2020amo, Roy:2023gzi, Maiti:2024dbl}, hybrid stars~\cite{Ranea-Sandoval:2018bgu}, and scenarios involving dark matter~\cite{Das:2021dru, Thakur:2024btu, Dey:2024vsw}. Authors in Ref.~\cite{Kalita:2023rbz} studied the non-radial oscillations with $\Delta$ baryons, but without metric perturbations. To achieve complete precision, a full GR treatment is required ~\cite{Kunjipurayil:2022zah, Pradhan:2022vdf, Roy:2023gzi, Zhao:2022tcw, Barman:2024zuo, Pradhan:2023zmg}.

This study examines the non-radial oscillation modes of NSs, applying both Cowling as well as GR methodology, allowing for a more precise evaluation of the method discrepancies, with various matter compositions, including nucleonic stars with $\Delta$-admixed matter and hyperon stars containing $\Delta$ baryons. For the first time, the analysis considers these compositions in scenarios where a hadron-quark phase transition occurs within the star. While previous research has focused on radial oscillations in NSs along with exotic phases, such as dark matter and deconfined quark matter~\cite{Rather:2023dom, Rather:2024hmo, 10.1093/mnras/stac2622, kokkostas, https://doi.org/10.48550/arxiv.2205.02076, PhysRevD.101.063025, PhysRevD.98.083001, https://doi.org/10.48550/arxiv.2211.12808}, this work extends the exploration to non-radial modes under similar conditions. The paper is organized as follows: Sec~\ref{sec:NS} provides the description of the NS used in this study. Sec.~\ref{sec:eos} outlines the EoS for the DD-RMF model with $\Delta$ baryons, the quark matter EoS, and the construction of the hybrid EoS. Sec.~\ref{nsprop} discusses the Tolman-Oppenheimer-Volkoff~(TOV) equations governing NS structure. Sec.~\ref{non-radial} details the non-radial oscillation analysis within the full GR framework. Sec.~\ref{results} presents the EoS and stellar properties, such as the speed of sound and mass-radius profiles for various compositions, with and without phase transitions. Sec.~\ref{frequency} examines the $f$-mode frequency as a function of stellar properties in both the Cowling and GR frameworks. Sec.~\ref{GW} introduces empirical fits and universal relations between the $f$-mode frequency and other key parameters. Sec.~\ref{summary} provides our concluding remarks. The Cowling approximation description along with its $f$-mode calculations are discussed in the Appendix for comparison with the GR results discussed in the paper.

\section{Neutron Star description}
\label{sec:NS}

\subsection{Microphysics}\label{sec:eos}
\subsubsection{Hadronic matter}

In this study, we describe the hadronic matter inside neutron stars using a density-dependent relativistic mean-field (DD-RMF) approach. This model is known for accurately reproducing experimental properties of nuclear matter and remains consistent with astrophysical constraints~\cite{dutra2014, lourencco2019, malik2022}. The interaction framework considers nucleons and other hadrons interacting via the exchange of virtual mesons. Specifically, the DD-RMF model used here includes the scalar meson $\sigma$, the vector mesons $\omega$ and $\phi$ (with hidden strangeness), and the isovector-vector meson $\vec{\rho}$.

The Lagrangian density serves as the foundational ansatz in any RMF theory, incorporating contributions from free baryons and mesons as well as interaction terms between them.
In the mean-field approximation, the Lagrangian of the relativistic model used here to describe hadronic interactions is given by
\begin{align}
\mathcal{L}_{\rm RMF} &= \sum_{b\in H}  \bar \psi_b \Big[  i \gamma^\mu\partial_\mu - \gamma^0  \big(g_{\omega b} \omega_0 + g_{\phi b} \phi_0+ g_{\rho b} I_{3b} \rho_{03}  \big)
\nonumber\\     
&- \left( m_b- g_{\sigma b} \sigma_0 \right)  \Big] \psi_b
- \frac{i}{2}\sum_{b\in \Delta}\bar{\psi}_{b\mu}\Big[\varepsilon ^{\mu \nu \rho \lambda }\gamma_5\gamma _\nu  \partial_\rho 
\nonumber\\
&- \gamma^0\left(g_{\omega b}\omega_0+g_{\rho b} I_{3b} \rho_{03} \right) -\left(m_b-g_{\sigma b}\sigma_0 \right)\varsigma ^{\mu \lambda }\Big]\psi_{b\nu}
\nonumber\\
&+\sum_\lambda\bar{\psi}_\lambda\left(i\gamma^\mu\partial_\mu-m_\lambda\right)\psi_\lambda- \frac{1}{2} m_\sigma^2 \sigma_0^2  +\frac{1}{2} m_\omega^2 \omega_0^2
\nonumber\\
&+\frac{1}{2} m_\phi^2 \phi_0^2 +\frac{1}{2} m_\rho^2 \rho_{03}^2,
\label{lagrangian}
\end{align}
where the first sum represents the Dirac-type interacting Lagrangian for the spin-1/2 baryon octet ($H=\{n,p,\Lambda,\Sigma^-,\Sigma^0,\Sigma^+,\Xi^-,\Xi^0\}$) and the second sum represents the Rarita-Schwinger interacting Lagrangian for the particles of the spin-3/2 baryon decuplet ($\Delta=\Delta^-,\Delta^0,\Delta^+,\Delta^{++}\}$), where $\varepsilon ^{\mu \nu \rho \lambda}$ is the Levi-Cicita symbol, $\gamma_5=i\gamma _0\gamma _1\gamma _2\gamma _3$ and $\varsigma ^{\mu \lambda }=\frac{i}{2}\left [ \gamma ^\mu,\gamma ^\lambda  \right ]$.
We note that spin-3/2 baryons are described by the Rarita-Schwinger Lagrangian density, where their vector-valued spinor has additional components compared to the four components in spin-1/2 Dirac spinors. However, as shown in~\cite{DePaoli}, the equations of motion for spin-3/2 particles can be simplified and written in a form analogous to those for spin-1/2 particles within the RMF framework.
The last sum describes the leptons admixed in the hadronic matter as a free non-interacting fermion gas ($\lambda=\{e,\mu\}$), as their inclusion is necessary in order to ensure the $\beta$-equilibrium and charge neutrality essential to stellar matter. The remaining terms account for the purely mesonic part of the Lagrangian.

\begin{table}[!ht]
\begin{center}
\caption{DDME2 parameters (top) and its predictions to the nuclear matter at saturation density (bottom).\label{T1} }
\begin{tabular}{ c c c c c c c }
\hline
$i$ & $m_i(\text{MeV})$ & $a_i$ & $b_i$ & $c_i$ & $d_i$ & $g_{i N} (n_0)$\\
 \hline
 $\sigma$ & 550.1238 & 1.3881 & 1.0943 & 1.7057 & 0.4421 & 10.5396 \\  
 $\omega$ & 783 & 1.3892 & 0.9240 & 1.4620 & 0.4775 & 13.0189  \\
 $\rho$ & 763 & 0.5647 & --- & --- & --- & 7.3672 \\
 \hline
\end{tabular}

\vspace{10pt}

\begin{tabular}{c|cc}
\hline 
Quantity & Constraints~\cite{dutra2014, Oertel:2016bki} & This model\\\hline
$n_0$ ($fm^{-3}$) & 0.148--0.170 & 0.152 \\
 $-B/A$ (MeV) & 15.8--16.5  & 16.14  \\ 
$K_0$ (MeV)& 220--260   &  252  \\
 $S_0$ (MeV) & 31.2--35.0 &  32.3  \\
$L_0$ (MeV) & 38--67 & 51\\
\hline
\end{tabular}
\label{tab:T1}
\end{center}
\end{table}

In DD-RMF models, coupling constants are typically functions of either the scalar density $n_s$ or the vector density $n_B$. Most commonly, vector density parameterizations are used, as they influence only the self-energy rather than the total energy~\cite{PhysRevLett.68.3408}. Here, we adopt the DD-RMF parametrization known as DDME2~\cite{ddme2}, where meson couplings scale with the baryonic density factor $\eta =n_B/n_0$ obeying the function
\begin{equation}
    g_{i b} (n_B) = g_{ib} (n_0) \frac{a_i +b_i (\eta+d_i)^2}{a_i +c_i (\eta+d_i)^2} 
\end{equation}
for $i=\sigma, \omega, \phi$ and 
\begin{equation}
    g_{\rho b} (n_B) = g_{ib} (n_0) \exp\left[ - a_\rho \big( \eta -1 \big) \right],
\end{equation}
for $i=\rho$. 

The model parameters are fitted to binding energies, charge radii, and differences between neutron and proton radii of spherical nuclei, as well as some bulk parameters related to infinite and pure nucleonic matter at $n_0$, namely, the saturation density itself, binding energy ($B/A$), incompressibility ($K_0$), and symmetry energy ($S_0$). All of them are shown in Table~\ref{tab:T1}, along with the value of the symmetry energy slope at $n_0$ ($L_0$). In order to determine the meson couplings to other hadronic species, we define the ratio of the baryon coupling to the nucleon one as $\chi_{ib}=g_{i b}/g_{i N}$, with $i = \{\sigma,\omega,\phi,\rho\}$. In this work, we consider hyperons and/or deltas admixed in the nucleonic matter and follow the proposal of~\cite{Lopes1} to determine their respective $\chi_{ib}$ ratios. This calibration follows a unified approach based on symmetry principles, particularly the requirement that the Yukawa coupling terms in the Lagrangian density of DD-RMF models remain invariant under SU(3) and SU(6) group transformations. Hence, the couplings can be fixed to reproduce the potentials  $U_\Lambda =-28$~MeV, $U_\Sigma= 30$~MeV, $U_\Xi=-4$~MeV and $U_\Delta=-98$~MeV in terms of a single free parameter $\alpha_V$. Our choice of $\alpha_V=1.0$ for the baryon-meson coupling scheme corresponds to an unbroken SU(6) symmetry, and the values of $\chi_{ib}$ are shown in Table~\ref{T2} taking into account the isospin projections in the Lagrangian terms~\cite{issifu}.
\begin{table}[!ht]
\centering
\caption {Baryon-meson coupling constants  $\chi_{ib}$~\cite{Lopes1}.
\label{T2}}
\begin{tabular}{ c c c c c } 
\hline
 b & $\chi_{\omega b}$ & $\chi_{\sigma b}$ & $I_{3b}\chi_{\rho b}$ & $\chi_{\phi b}$  \\
 \hline
 $\Lambda$ & 2/3 & 0.611 & 0 & 0.471  \\  
  $\Sigma^{-}$,$\Sigma^0$, $\Sigma^{+}$ & 2/3 & 0.467 & $-1$, 0, 1 & -0.471 \\
$\Xi^-$, $\Xi^0$  & 1/3 & 0.284 & $-1/2$, 1/2 & -0.314 \\
  $\Delta^-$, $\Delta^0$, $\Delta^+$, $\Delta^{++}$   & 1 & 1.053 & $-3/2$, $-1/2$, 1/2, 3/2 & 0  \\
  \hline
\end{tabular}
\end{table}

From the Lagrangian \eqref{lagrangian}, thermodynamic quantities can be calculated in the standard way for RMF models. The baryonic and scalar densities of a baryon of the species $b$ are given, respectively, by
\begin{equation}
n_b = \frac{\lambda_b}{2\pi ^{2}}\int_{0}^{{k_F}_b}dk\, k^{2}=\frac{\lambda_b}{6\pi ^{2}}{k_F}_b^{3}, \label{eq:rhobarion}
\end{equation} 
and
\begin{equation}
 n^s_b=\frac{\lambda_b}{2\pi ^{2}}\int_{0}^{{k_F}_b} dk \frac{k^{2}m_b^\ast}{\sqrt{k^{2}+{m_b^\ast}^{2}}}, \label{eq:rhoscalar}
\end{equation} 
with ${k_F}$ denoting the Fermi momentum,  since we assume the stellar matter to be at zero temperature, and $\lambda_b$ is the spin degeneracy factor (2 for the baryon octet and 4 for the delta resonances).
The effective masses are 
\begin{equation}
    m_b^\ast =m_b- g_{\sigma b} \sigma_0 .
\end{equation}

The energy density is given by
\begin{align}\label{1a}
\varepsilon_B &= \sum_b \frac{\gamma_b}{2\pi^2}\int_0^{{k_{F}}_b} dk k^2 \sqrt{k^2+{m_b^\ast}^2}
\nonumber\\
&+ \sum_\lambda \frac{1}{\pi^2}\int_0^{{k_{F}}_\lambda} dk k^2 \sqrt{k^2+m_\lambda^{2}}
\nonumber\\
&+ \frac{m_\sigma^2}{2} \sigma_0^2+\frac{m_\omega^2}{2} \omega_0^2 +\frac{m_\phi^2}{2} \phi_0^2 +\frac{m_\rho^2}{2} \rho_{03}^2 .
\end{align}
The effective chemical potentials read
\begin{equation}
      \mu_b^\ast = \mu_b- g_{\omega b} \omega_0 - g_{\rho b} I_{3b} \rho_{03} - g_{\phi b} \phi_0 - \Sigma^r,
\end{equation}
where $\Sigma^r$ is the rearrangement term, necessary to ensure thermodynamical consistency due to the density-dependent couplings,
\begin{align}\label{ret}
    \Sigma^r ={}& \sum_b \Bigg[ \frac{\partial g_{\omega b}}{\partial n_b} \omega_0 n_b+\frac{\partial g_{\rho b}}{\partial n_b} \rho_{03} I_{3b}  n_b+ \frac{\partial g_{\phi b}}{\partial n_b} \phi_0 n_b \nonumber \\
    &- \frac{\partial g_{\sigma b}}{\partial n_b} \sigma_0 n_b^s\Bigg],
\end{align}
and the $\mu_b$ are determined by the chemical equilibrium condition
\begin{equation}
    \mu_b=\mu_n-q_b\mu_e, \label{beta}
\end{equation}
in terms of the chemical potential of the neutron and the electron, with $\mu_\mu=\mu_e$. The particle populations of each individual species are determined by Eq.~\eqref{beta} together with the charge neutrality condition $\sum_i n_iq_i=0$, where $q_i$ is the charge of the baryon or lepton $i$.
The pressure, finally, is given by
\begin{equation}
    P =\sum_i \mu_i n_i - \epsilon+n_B \Sigma^r,
\end{equation}
which receives a correction from the rearrangement term to guarantee thermodynamic consistency and energy-momentum conservation~\cite{Typel1999, PhysRevC.52.3043}. In the above expression, $\epsilon$ is the total energy density including leptons.

\subsubsection{Deconfined quark matter}

In this study, we adopt the density-dependent quark mass (DDQM) model~\cite{Xia2014} to describe quark matter, a simple and versatile framework well-suited for investigating the deconfinement phase transition in hybrid stars~\cite{backes2021effects}. The DDQM model simulates the QCD quark confinement through density-dependent quark masses defined by
\begin{equation}
    m_i = m_{i0}+\frac{D}{n_B^{1/3}}+Cn_B^{1/3} = m_{i0}+m_I,
    \label{masses}
\end{equation}
where ${m_{i0}}$ (${i = u, d, s}$) is the current mass of the $i$th quark, ${n_B}$ is the baryon number density and ${m_I}$ is the density-dependent term that encompasses the interaction between quarks. This model-free parameters $C$ and $D$ dictate linear confinement and the leading-order perturbative interactions, respectively~\cite{Xia2014}.

Introducing density dependence for state variables, such as density, temperature, or magnetic field, requires careful handling to maintain thermodynamic consistency, analogous to the approach in Eq. \eqref{ret} for the DD-RMF model. We follow the formalism in~\cite{Xia2014}, which ensures thermodynamic consistency in DDQM. At zero temperature, the fundamental differential relation for energy density reads
\begin{equation}
    \text{d}\varepsilon = \sum_i \mu_i\text{d}n_i,
    \label{diff-fundamental-eq}
\end{equation}
where ${\varepsilon}$ is the matter contribution to the energy density of the system, ${\mu_i}$ are the particle chemical potentials and ${n_i}$ are the particle densities. 

To express this model in terms of effective chemical potentials, we represent the energy density as for a free system as
\begin{equation}
    \varepsilon = \Omega_0 (\{\mu_i^*\},\{m_i\})+\sum_i \mu_i^* n_i,
    \label{free-system-fundamental-eq}
\end{equation}
 using density-dependent quark masses ${m_i(n_B)}$  and effective chemical potentials ${\mu_i^*}$, where ${\Omega_0}$ is the thermodynamic potential of a free system. We can differentiate this form to yield
\begin{equation}
    \text{d}\varepsilon = \text{d}\Omega_0+\sum_i \mu_i^* \text{d} n_i+\sum_i n_i \text{d}\mu_i^*.
    \label{initial-diff-free-system}
\end{equation}
Explicitly, we can write d${\Omega_0}$ as
\begin{equation}
    \text{d} \Omega_0 = \sum_i \frac{\partial \Omega_0}{\partial \mu_i^*} \text{d}\mu_i^*+\sum_i \frac{\partial \Omega_0}{\partial m_i}\text{d}m_i,
\end{equation}
with
\begin{equation}
    \text{d}m_i = \sum_j \frac{\partial m_i}{\partial n_j} \text{d}n_j,
\end{equation}
where, to ensure thermodynamic consistency, the densities are connected to the effective chemical potentials by
\begin{equation}
    n_i = -\frac{\partial \Omega_0}{\partial \mu_i^*}.
\end{equation}
Eq. \eqref{initial-diff-free-system} can then be rewritten as
\begin{equation}
    \text{d}\varepsilon = \sum_i \left(\mu_i^*+\sum_j \frac{\partial \Omega_0}{\partial m_j} \frac{\partial m_j}{\partial n_i}\right) \text{d}n_i,
    \label{diff-free-system}
\end{equation}
providing a relation between the real and effective chemical potentials,
\begin{equation}
    \mu_i = \mu_i^*+\sum_j \frac{\partial \Omega_0}{\partial m_j} \frac{\partial m_j}{\partial n_i}.
\end{equation}
Consequently, from the fundamental relation $P = -\varepsilon+\sum_i \mu_i n_i$, the pressure ${P}$ is given by
\begin{equation}
      P ={}-\Omega_0+\sum_{i,j} \frac{\partial \Omega_0}{\partial m_j}n_i\frac{\partial m_j}{\partial n_i},
    \label{pressure-quarks}
\end{equation}
yielding a thermodynamically consistent EoS for quark matter.

The EoS for the quark model is derived using experimentally consistent quark masses and selected parameters suited for hybrid stars based on phase coexistence with various hadronic configurations. The transition point between phases is highly sensitive to the free parameters of the DDQM model, which lacks strong empirical constraints. Therefore, parameter selection often involves considering the stability window under the Bodmer-Witten hypothesis~\cite{Bodmer1971, Witten:1984}, which posits that strange quark matter -- comprising roughly equal amounts of $u$, $d$, and $s$ quarks -- could be more stable than hadronic matter. If true, neutron stars could convert entirely into strange stars. However, since our focus is on hybrid stars, we exclude parameter sets that satisfy this hypothesis. Additionally, studies have shown that for high values of the $C$ parameter, the surface density of strange stars can approach or fall below nuclear saturation density, indicating a possible phase transition. Such parameters also result in hybrid star phase transitions at densities above nuclear saturation and yield strange stars with masses around 2 $M_\odot$.
Ref.~\cite{Backes_2021} provides a detailed analysis of how DDQM parameters affect strange matter stability, and the specific choice of the quark matter-free parameters $C$ and $D$ adopted here is discussed in detail in Ref.~\cite{Rather:2024hmo}.

\subsubsection{Phase transition and hybrid EoS construction}

Studying matter under extreme conditions is inherently difficult due to the complexity of QCD. The two main theoretical approaches -- lattice QCD (LQCD) and effective models -- each have significant limitations. LQCD faces challenges such as the sign problem, computational constraints, and limited applicability at high chemical potentials, making it ineffective for mapping the QCD phase diagram in these regimes (see~\cite{Nagata:2021ugx}). Consequently, effective models are often employed, particularly in the context of compact objects like neutron stars.

A longstanding tension exists between LQCD and effective models regarding the nature of the QCD phase transition. LQCD indicates a smooth crossover around 160–170 MeV at low chemical potentials~\cite{Aoki:2006we, Luo:2020pef}, while effective models predict a first-order transition at high densities. This transition is expected to culminate in a critical endpoint (CEP), beyond which it becomes second-order. However, the existence and precise location of the CEP remain uncertain~\cite{Bazavov:2017dus, HotQCD:2017qwq}. For example,~\cite{Fukushima:2010bq} suggests that at zero temperature, the transition onset requires a chemical potential exceeding 1050 MeV in the Polyakov loop formalism.

The characteristics of the transition vary according to the quark and hadron EoS models employed. In this study, we assume that the hadron-quark deconfinement transition is a first-order phase transition, as predicted by effective models in the high-density region of the QCD phase diagram. A phase transition can occur as either a Maxwell or a mixed phase (also called Gibbs)  transition. In a Maxwell transition, the phases remain separate and maintain local charge conservation, whereas, in a mixed transition, quarks and hadrons coexist over a range of baryonic densities with global charge conservation. The hadron-quark phase surface tension serves as the primary criterion for determining the type of phase transition. Values above 60 MeV/fm$^2$ favor a Maxwell transition~\cite{Voskresensky:2002hu, Maruyama:2007ey}, while lower values suggest a mixed transition. Given the uncertainties in surface tension estimates~\cite{Pinto:2012aq, Lugones:2013ema, Lugones:2016ytl, Lugones:2018qgu}.  The thermodynamic description of this process involves matching the EoS of the two phases and identifying the point of phase coexistence.

In this study, we apply the Maxwell construction, producing a hybrid EoS with a first-order phase transition at critical values of baryonic chemical potential and pressure. According to Gibbs' criteria, the transition occurs at the point where
\begin{align}
     P^{(i)}={}&P^{(f)}=P_0,\\
  \mu^{(i)}(P_0)={}&\mu^{(f)}(P_0)=\mu_0,
\label{eq:gibbscon}
\end{align}
sets the transition between the initial (${i}$) and final (${f}$) homogeneous phases, both at ${T=0}$ MeV, with
\begin{equation}
 \mu^{ (i,f)}=\frac{\varepsilon^{(i,f)}+P^{(i,f)}}{n_B^{(i,f)}},
\end{equation}
where ${\varepsilon^{(i,f)}}$, ${P^{(i,f)}}$ and ${n_B^{(i,f)}}$ are the total energy density, pressure, and baryon number density, obtained from the EoS of each phase.
The conditions above the values of $P_0$ and $\mu_0$ are to be determined from the equations of state of both hadronic and deconfined quark phases. 
The transition point location, for a given baryonic composition in the hadronic phase, will be notably influenced by the choice of the free parameters for the DDQM model~\cite{backes2021effects}.

\subsection{Macrophysics}
\label{nsprop}

Moving from micro to macrophysics involves applying the EoS for the dense matter to conditions of mechanical (or hydrostatic) equilibrium, as NS is assumed to have stable internal structures.
The intense gravitational field of NS makes their structure and dynamical evolution be governed by Einstein's equations of General Relativity,
\begin{equation}
    G_{\mu \nu} = R_{\mu \nu} - \frac{1}{2} R g_{\mu \nu} = 8\pi T_{\mu \nu}, 
\end{equation}
where $R_{\mu \nu}$ is the Ricci tensor and $R$ is the Ricci scalar, and $T_{\mu \nu}$ is the energy-momentum tensor.

One can obtain the TOV equations~\cite{PhysRev.55.364, PhysRev.55.374} for the equilibrium structure of NSs by solving the Einstein field equation with the below-defined metric, 
\begin{align}\label{tov1}
\frac{dP(r)}{dr}={}& -\frac{[\varepsilon(r) +P(r)][m(r)+4\pi r^3 P(r)]}{r^2(1-2m(r)/r) } ,\\\label{tov2}
\frac{dm(r)}{dr}={}& 4\pi r^2 \varepsilon(r) ,
\end{align}
by taking the $T_{\mu \nu}$ of an homogeneous fluid,
\begin{equation}
    T_{\mu \nu} = P g_{\mu \nu}+(P+\varepsilon) u_{\mu}u_{\nu},
\end{equation}
where $g_{\mu \nu}$ is the metric tensor, $P$ is the pressure, $\varepsilon$ is the energy density, and $u_{\mu}$ is the four-velocity, and considering static spherically symmetric stars, described by the Schwarzschild metric as~\cite{1977ApJ...217..799C}
\begin{equation}\label{metric}
    ds^{2} = e^{\nu(r)} dt^{2} -e^{\lambda(r)} dr^{2} - r^{2} (d\theta^{2}+\sin^{2} \theta d\phi ^{2}),
\end{equation}
where $e^{\nu(r)}$ and $e^{\lambda(r)}$ are the metric functions. 

Using the given EoS, the TOV Eqs. \eqref{tov1}-\eqref{tov2} are solved with initial conditions $m(r=0) = 0$ and $P(r=0) = P_c$, where $P_c$ represents the central pressure. The star's radius, $R$, is defined where the pressure vanishes at the surface, $P(R) = 0$, and the total mass is then given by $M = m(R)$.

\section{Oscillation modes}
\label{non-radial}

\subsection{Non-radial oscillations in general relativity}
\label{sec:GRR}

To determine the frequencies of the $f$-modes in the full general relativity formalism, we solve Einstein's field equations assuming that the gravitational waves represent perturbations to the static background spacetime metric of a non-rotating neutron star. The perturbed metric is given by
\begin{equation}
g_{\mu \nu} = g^{0}_{\mu \nu}+h_{\mu \nu},
\end{equation}
Only even-parity perturbations of the Regge-Wheeler metric are significant in this context~\cite{PhysRevLett.24.737}
A small perturbation, \( h_{\mu \nu} \), is introduced to a static, spherically symmetric background metric, which is described as:
\begin{align}
ds^2 ={}& - e^{\nu(r)} [1+r^l H_0(r)e^{i\omega t} Y_{lm}(\phi,\theta)]c^2 dt^2\nonumber\\&
+e^{\lambda(r)} [1-r^l H_0(r)e^{i\omega t} Y_{lm}(\phi,\theta)]dr^2 \nonumber\\&
+ [1-r^l K(r)e^{i\omega t}Y_{lm}(\phi,\theta)]r^2 d\Omega^2\nonumber\\&
-2i\omega r^{l+1}H_1(r)e^{i\omega t}Y_{lm}(\phi,\theta) dt~dr,
\end{align}
where, \( H_0 \), \( H_1 \), and \( K \) represent the radial perturbations of the metric, while the angular dependence is captured by the spherical harmonics \( Y_l^m \). The time dependence of the perturbed metric components can be expressed using the factor \( e^{i\omega t} \) for a wave mode. Here \( \omega \) is a complex quantity, as the waves decay due to the imposed open boundary conditions. The real part of \( \omega \) represents the oscillation frequency, while the imaginary part corresponds to the inverse of the wave mode's gravitational wave damping time (positive).

The perturbations of the energy-momentum tensor of the fluid must also be considered in the Einstein equations. The components of the Lagrangian displacement vector \( \xi^a(r, \theta, \phi) \) describe the perturbations of the fluid within the star:
\begin{align}
\xi^r ={}& r^{l-1}e^{-\frac{\lambda}{2}}W Y^l_m e^{i\omega t} ,\label{eq:xi_radial} \nonumber \\
\xi^\theta ={}& -r^{l-2} V \partial_\theta Y_m^l e^{i\omega t}, \nonumber  \\
\xi^\phi ={}& -\frac{r^{l-2}}{ \sin^{2}\theta} V\partial_\phi Y_m^l e^{i\omega t} .
\end{align}
here, $W$ and $V$ are functions of $r$ that represent fluid perturbations confined to the star's interior.

The gravitational wave equations can then be written as a set of four coupled linear differential equations for the four perturbation functions, $H_1$, $K$, $W$, and $X$, which do not diverge inside the star for any given value of $\omega$.~\cite{Lindblom:1983ps, Detweiler:1985zz},
\begin{align}
r\frac{dH_1}{dr}&=-[l+1+2b e^\lambda+4\pi r^2 e^\lambda(p-\varepsilon)] H_1
\nonumber\\
&+ e^\lambda[H_0+K-16\pi(\varepsilon+p)V]\,, 
\label{eq:ODE_DL1} \\
r\frac{dK}{dr}&= H_0+(n_l+1)H_1 
\nonumber \\
&+[e^\lambda \textrm{Q}-l-1]K-8\pi(\varepsilon+p)e^{\lambda/2}W \,, 
\label{eq:ODE_DL2}\\
r\frac{dW}{dr}&=-(l+1)[W+le^\frac{\lambda}{2}V] 
\nonumber \\
&+r^2 e^{\lambda/2}\left[\frac{e^{-\nu/2}X}{ (\varepsilon+p) c_{\rm ad}^2}+\frac{H_0}{2}+K\right], 
\label{eq:ODE_DL3}
\end{align}
\begin{align}
&r\frac{dX}{dr}= -lX+\frac{(\varepsilon+p)e^{\nu/2}}{2}\times
\nonumber \\
&\times\Bigg\{ (1-e^\lambda \textrm{Q})H_0+(r^2\omega^2e^{-\nu}+n_l+1)H_1 
\nonumber  \\
&+(3e^\lambda\textrm{Q}-1)K
-\frac{4(n_l+1)e^\lambda\textrm{Q}}{r^2}V 
\nonumber\\
&-2\left[\omega^2 e^{\lambda/2-\nu} +4\pi(\varepsilon+p)e^{\lambda/2} -r^2\frac{d}{dr} \left(\frac{e^{{\lambda/2}} \textrm{Q}}{r^3} \right)\right]W \Bigg\} \,, \label{eq:ODE_DL4}
\end{align}

where,
\begin{align}
Q(r) = b(r) + \frac{4\pi G r^2 p(r)}{c^4} f.
\end{align}

Here, $b(r) = 2Gm(r)/(c^2 r)$, with $m(r)$ and $p(r)$ denoting the enclosed mass and pressure at radius $r$, respectively. The number of fluid perturbation variables is given by $n_l = (l - 1)(l + 2)/2$. The angular dependence is described by the spherical harmonics $Y_{lm}$, characterized by the angular quantum number $l$ and azimuthal quantum number $m$; the latter is degenerate for the nonrotating neutron stars considered here. In this work, the adiabatic sound speed, \( c_{\rm ad}^2 \), which characterizes oscillations in neutron star matter, is approximated by the equilibrium sound speed defined as $c_{\rm eq}^2 = dp/d\varepsilon$~\cite{Kunjipurayil:2022zah,Zhao:2022tcw}. The behavior of \( c_{\rm eq}^2 \) is illustrated in \Cref{figcs}.

Perturbations at the center of the star $r=0$ are subject to
the boundary conditions  $X(R)=0$, $W(0)=1$,
\begin{align}
&X(0) = (\varepsilon_0+p_0)e^{\nu_0/2}\times
\nonumber \\
&\times \bigg\{ \left[ \frac{4\pi}{3}(\varepsilon_0+3p_0) 
- \frac{\omega^2}{l} e^{-\nu_0} \right]W(0)+\frac{K(0)}{2} \bigg\},
\label{eq:boundary_conditions_1}
\end{align}
and
\begin{align}
H_1(0) = \frac{lK(0)+8\pi(\varepsilon_0+p_0)W(0)}{n_l+1}.
\label{eq:boundary_conditions}
\end{align}

The final boundary condition is derived by solving two trial solutions with 
\( K(0) = \pm(\varepsilon_0+p_0) \) and then forming a linear combination to satisfy the condition 
\( X(r=R) = 0 \), which ensures there are no pressure variations at the surface. By design, 
\( H_0(0) = K(0) \).

At the star's surface, small arbitrary values are assigned to the functions 
\( H_1 \), \( K \), and \( W \), and backward integration is performed until reaching the point where forward 
integration from the star’s center ends. The forward and backward solutions are then matched at this point. 
The quasinormal mode frequency for the star is determined by solving the Zerilli equation,
\begin{equation}
\frac{d^2Z}{dr^{*2}}=[V_Z(r)-\omega^2]Z \,. \label{eq:zerilli}
\end{equation}
The Zerilli function, as expressed in Eq.~(20) of~\cite{Kunjipurayil:2022zah}, depends solely on the perturbation variables \(H_1\) and \(K\), since the fluid perturbations \(W\), \(V\), and \(X\) vanish outside the star. The value \(Z(r)\) at the star's surface is determined using the values of \(H_1\) and \(K\) at the surface. Beyond the star, Eq. (\ref{eq:zerilli}) is numerically integrated starting from the surface and extending outward to a distance corresponding to \(r = 25~\omega^{-1}\)~\cite{Kunjipurayil:2022zah}. The value of \(Z\) at \(r = 25~\omega^{-1}\) is matched with the corresponding value obtained from the asymptotic expansion of \(Z\), which is valid far from the neutron star's surface. To account for the imaginary component of $\omega$, which is over a thousand times smaller than its real counterpart, it is essential to maintain a relative error of $10^{-6}$ in our ODE solver for the variables $H_1$, $K$, $W$, $X$, and $Z$.


\section{Numerical results and discussion}
\label{results}
\subsection{Equation of State and Mass-Radius relations}
\label{mr}

\begin{figure*}
\begin{minipage}{0.47\textwidth}		 		
\includegraphics[width=\textwidth]{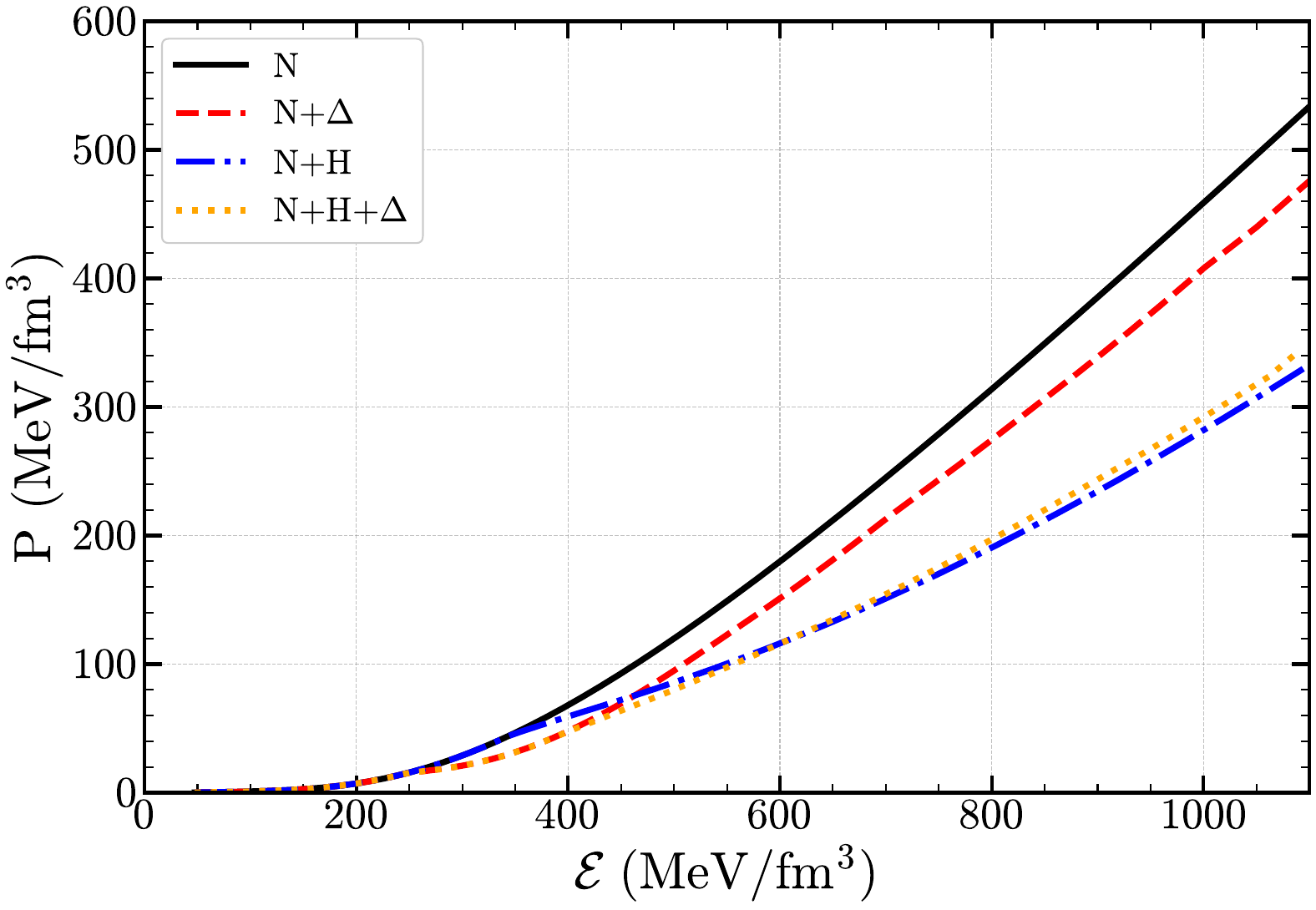}
\end{minipage}
\begin{minipage}{0.47\textwidth}
\includegraphics[width=\textwidth]{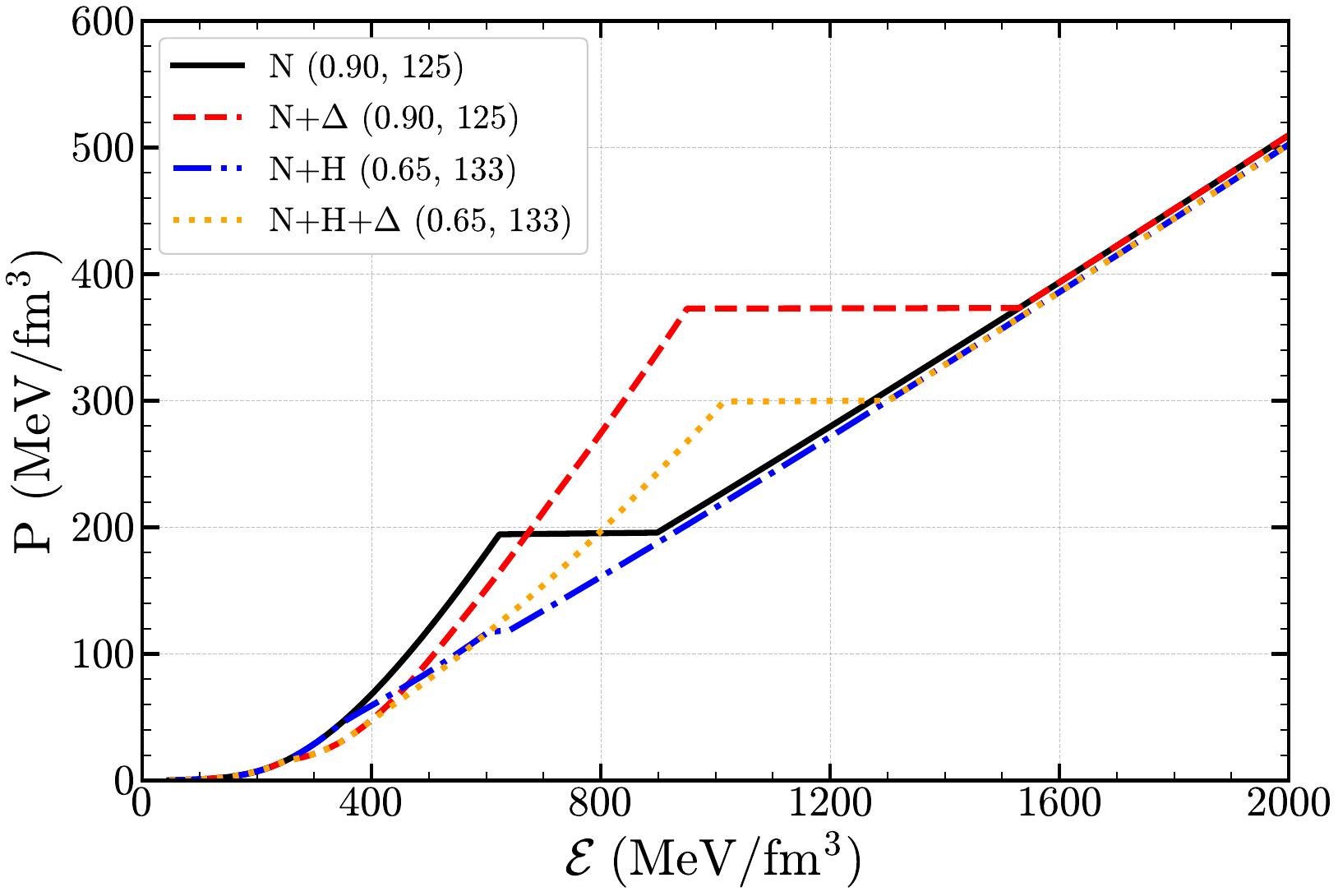}
\end{minipage}
\caption{Energy density and pressure variation for the given DD-ME2 parameter set without (left) and with (right)  phase transition to the quark matter at different quark model parameters ($C, D^{1/2}$). The solid line represents the pure nucleonic matter (N) while dashed, dash-dotted, and dotted lines represent the EoS for $\Delta$-admixtured nuclear matter $\Delta$(N+$\Delta$), with hyperons (N+ H), and $\Delta$-admixtured hyperonic matter (N+H+$\Delta$), respectively.}
\label{figeos}	 	
\end{figure*}

Figure~\ref{figeos} illustrates how pressure varies with energy density (i.e., the EoS) for a neutron star under beta-equilibrium and charge-neutral conditions. The left panel shows different compositions of hadronic matter: pure nucleonic matter (N), $\Delta$-admixtured nuclear matter (N+$\Delta$), with hyperonic matter (N+H), and $\Delta$-admixtured hyperonic matter (N+H+$\Delta$), and the right panel shows the EoS when a phase transition to the quark matter is included. From the left plot, we can see that the pure nucleonic matter results in a stiffer EoS at high densities. The appearance of $\Delta$ particles softens the EoS, as additional particle types distribute the Fermi pressure across multiple degrees of freedom.  With only nucleons and hyperons present, the EoS softens further, but adding $\Delta$ particles to hyperonic matter (N+H+$\Delta$) introduces complexities. As seen in Figure~\ref{figeos}, N+H+$\Delta$ is softer than N+H at low densities but becomes stiffer as density increases. This stiffening occurs because the $\Delta^{-}$ baryon replaces a neutron-electron pair at the Fermi surface, which is energetically favorable due to an attractive potential. Neutral particles, as the $\Lambda$ and $\Delta^{0}$, appear later~\cite{PhysRevC.106.055801}. 

Regarding the phase transition, the presence of ${\Delta}$s causes a shift in the coexistence point towards higher densities for the same deconfined EoS, which is linked to the aforementioned effect. Post-phase transition, the EoS at higher densities is much more uniform compared to its hadronic counterpart. For instance, the parameter set (${C, D^{1/2}}$) = (0.90, 125 MeV) results in only a slightly stiffer EoS than (${C, D^{1/2}}$) = (0.65, 133 MeV). However, the position of the coexistence point plays the most crucial role when constructing the hybrid EoS. Thus, for hybrid N+$\Delta$ EoS, the phase transition takes place at a very high density compared to hybrid N+H+$\Delta$ EoS. For the hybrid N+H EoS, the hadron-quark phase transition region is small and occurs at low density compared to the others. This implies a large quark phase present in comparison to the other hybrid EoSs. 

\begin{figure*}[t]
		\begin{minipage}[t]{0.47\textwidth}		 		
  \includegraphics[width=\textwidth]{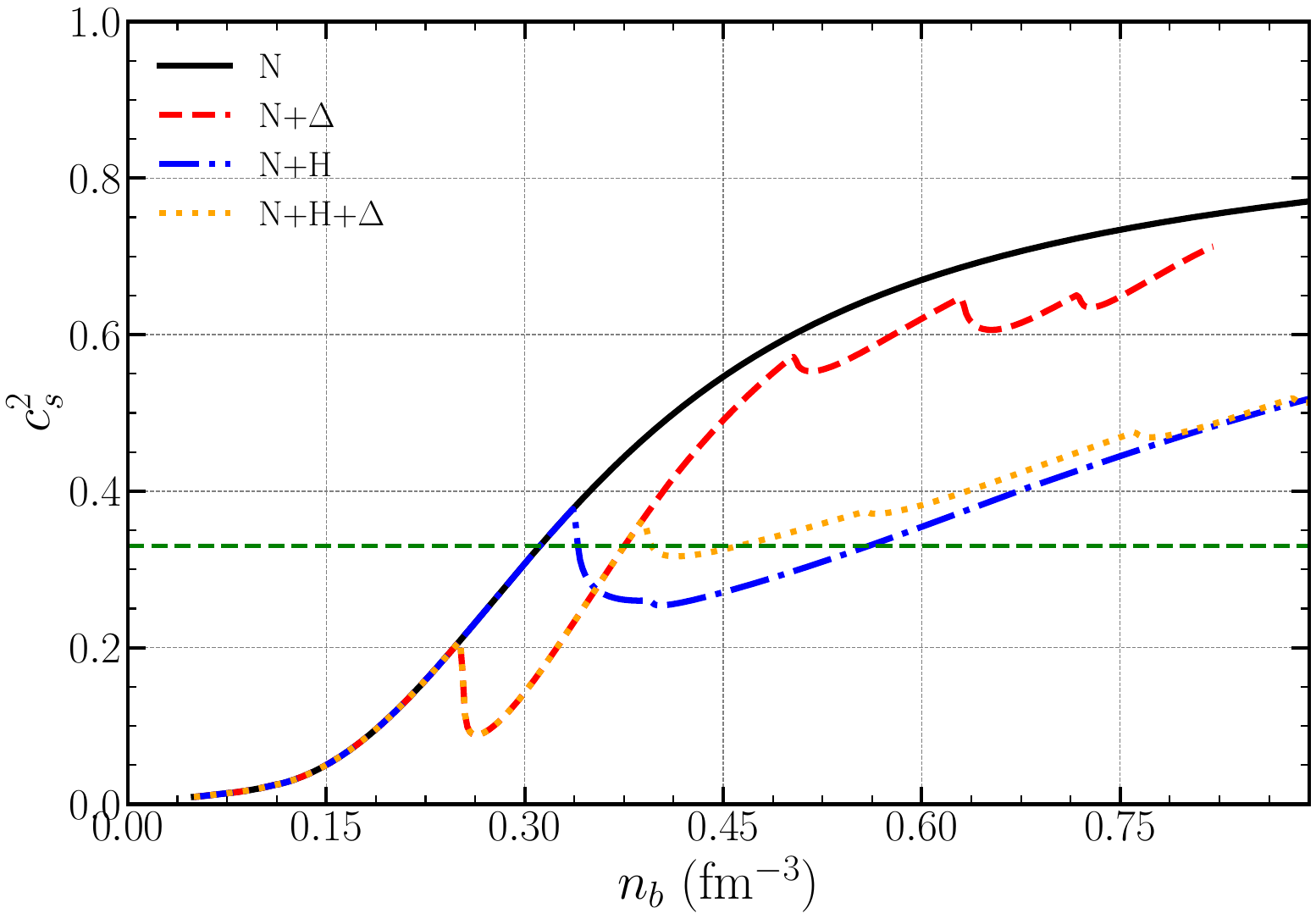}
			 	\end{minipage}
		 		\begin{minipage}[t]{0.47\textwidth}
			 		\includegraphics[width=\textwidth]{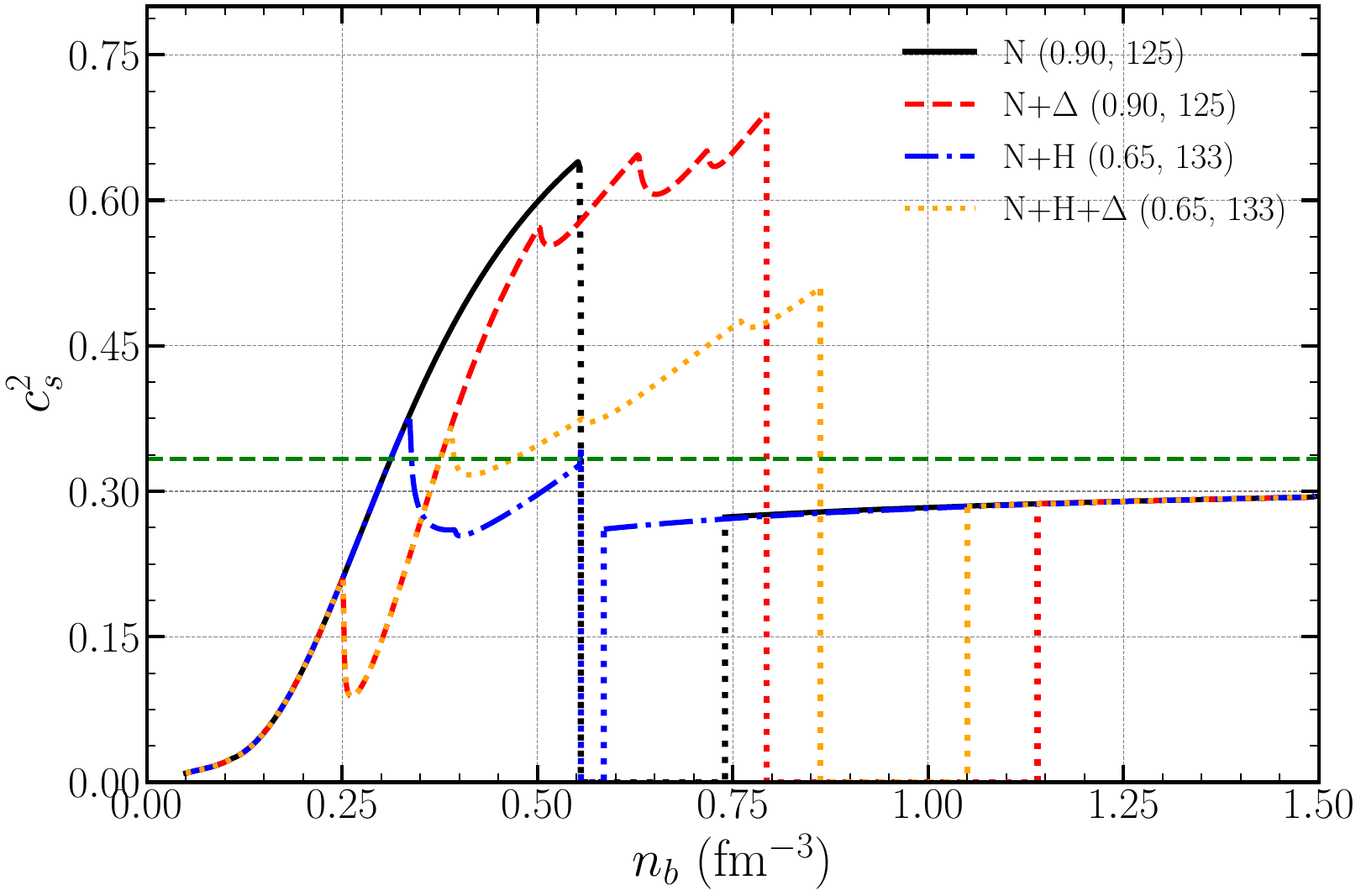}
			 	\end{minipage}
			 			\caption{Speed of sound squared as a function of number density for the different hadronic compositions of EoS without (left) and with phase transition (right) to the quark matter. The dotted lines in the right plot correspond to the mixed-phase region where $c_s^{2}$ drops to zero. The green dashed line in both plots represents the conformal limit $c_s^{2}$ = 1/3.}
		\label{figcs}	 	
     \end{figure*}
  
 Figure~\ref{figcs} depicts the behavior of squared speed of sound as a function of number density for different compositions of the matter studied in this work, without (left) and with (right) phase transition. Thermodynamic stability ensures that $c_s^{2}> 0$ and causality implies an absolute bound $c_s^2\leq 1$. For very high densities, perturbative QCD findings anticipate an upper limit of $c_s^{2} =1/3$~\cite{PhysRevLett.114.031103}. The two solar mass requirements, according to several studies~\cite{PhysRevLett.114.031103, PhysRevC.95.045801, Tews_2018}, necessitates a speed of sound squared that exceeds the conformal limit ($c_s^{2}$ = 1/3), revealing that the matter inside of NS is a highly interacting system.
In Figure~\ref{figcs}, the $c_s^{2}$ for pure nucleonic matter is significantly high, reaching a value of 0.75 at the maximum mass configuration. In the appearance of different particles, one can see the kinks corresponding to the
onset of a new particle species, resulting in noticeable changes at the onset of each type. Both pure nucleonic and $\Delta$-mixed nuclear matter exceed the conformal limit. Additionally, the N+H+$\Delta$ EoS shows a higher value of $c_s^{2}$ compared to N+H EoS at intermediate densities due to the early emergence of $\Delta^{-}$ particles. For the maximum mass configuration, the $c_s^{2}$ for N+H is 0.54 while for N+H+$\Delta$ is 0.51.

When transitioning to quark matter (right plot), $c_s^{2}$ exhibits a discontinuity as the density varies abruptly in the interface between the phases. For different particle combinations, kinks are observed before phase transitions, with hybrid N, N+$\Delta$, and N+H+$\Delta$ EoS violating the conformal limit at low densities. The N+H+$\Delta$ composition predicts a higher $c_s^{2}$ due to early $\Delta^{-}$ appearance and delayed quark transition. At high energy densities, all speed of sound values stays well below the conformal limit, unlike previous observations, due to the expected approach of a deconfined EoS towards the conformal limit from below~\cite{Bedaque:2014sqa}. For all the cases, the speed of sound at the maximum mass configuration lies within the range of 0.25-0.27 because of the transition to the quark matter.

\begin{figure*}[t]
\begin{minipage}[t]{0.49\textwidth}		 		
\includegraphics[width=\textwidth]{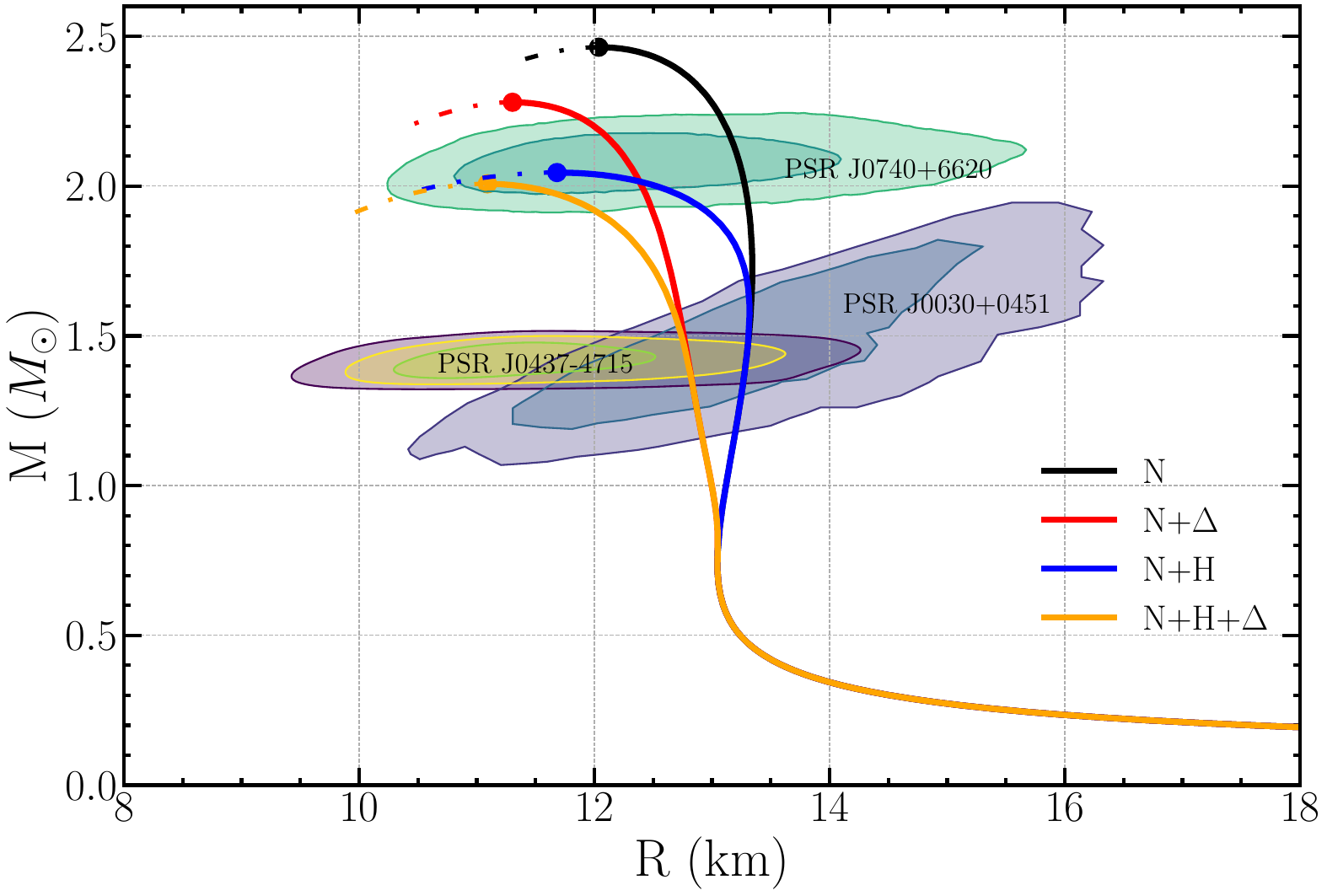}
\end{minipage}
\begin{minipage}[t]{0.47\textwidth}
\includegraphics[width=\textwidth]{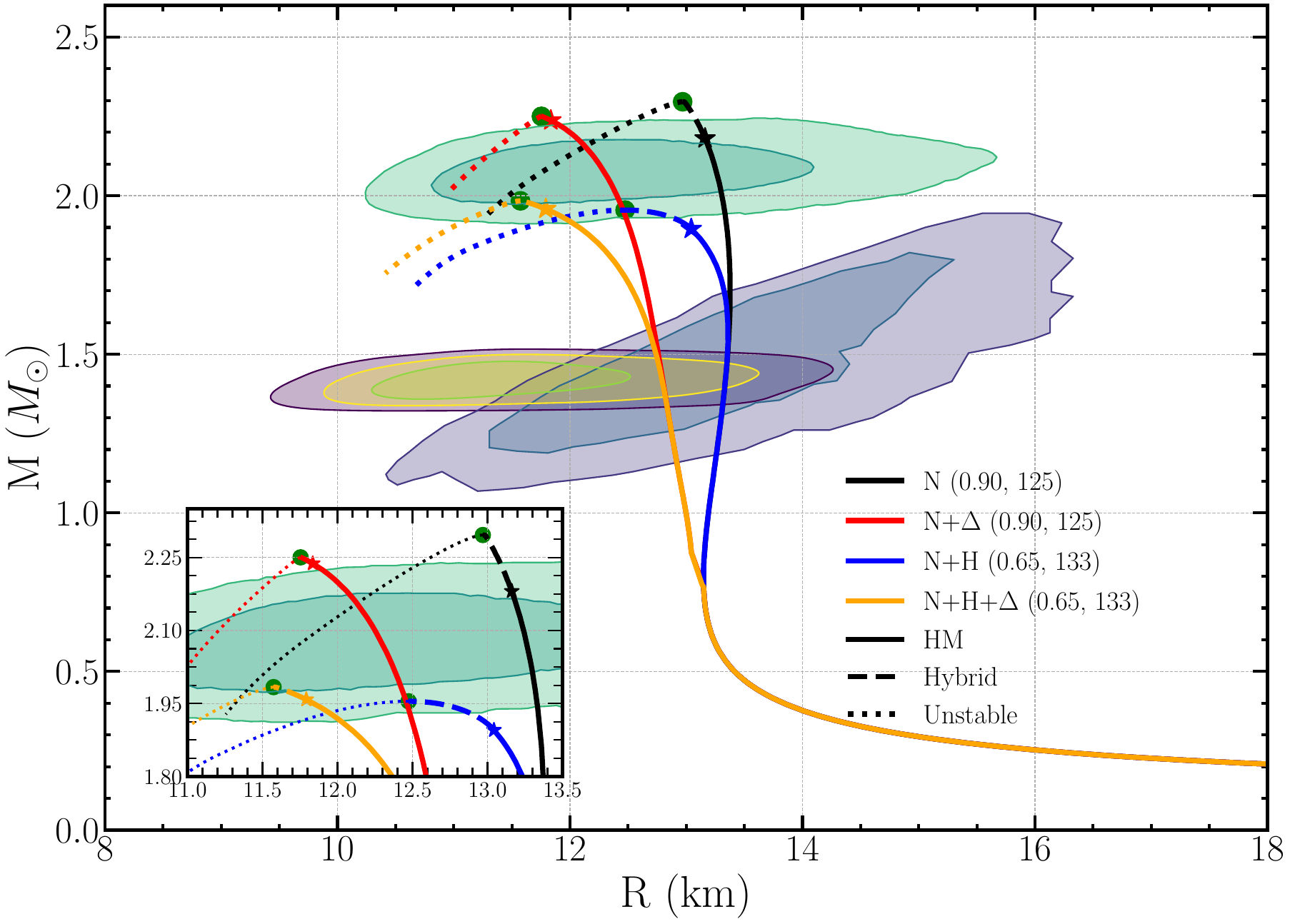}
\end{minipage}
\caption{Left: Mass-Radius relation for the EoS with different hadronic compositions. The solid lines represent the stable part with a solid dot marking the last stable point, hence the maximum mass configuration. The dash-dotted line represents the unstable part.  Right: Same as the left plot but with a phase transition to the quark matter. The solid lines represent the hadronic branch. The star symbol corresponds to the beginning of the hybrid stars branch. The solid dot represents the last stable point reached in the center of the maximum-mass solution of the TOV equation. The dotted line represents the unstable part. The various shaded areas are credibility regions from the mass and radius inferred from the analysis of PSR J0740+6620, PSR J0030+0451, and PSR J0437-4715~\cite{miller2021,2021ApJ...918L..27R, Miller_2019a, Riley_2019, Choudhury:2024xbk}. }
\label{figmr}	 	
\end{figure*}
     
Figure~\ref{figmr} illustrates the mass-radius relationship based on solutions of the TOV equations for various EoSs. The unified EoS employs the Baym-Pethick-Sutherland (BPS) EoS~\cite{Baym:1971pw} for the outer crust, while the inner crust EoS is generated using the DD-ME2 parameter set in the Thomas-Fermi approximation~\cite{PhysRevC.79.035804, PhysRevC.94.015808, rather2020effect}. The left plot represents the MR relations for different compositions of nuclear matter without a phase transition. In contrast, the right plot represents the hybrid EoS with the same compositions of nuclear matter but with a phase transition to the quark matter. From the left plot, for purely nucleonic matter, the maximum mass reaches 2.46\,$M_{\odot}$ with a radius of 12.04 km. When $\Delta$ baryons are included, both the maximum mass and corresponding radius decrease to 2.28\,$M_{\odot}$ and 11.30 km. The presence of hyperons softens the EoS, reducing the maximum mass to 2.04\,$M_{\odot}$ with a radius of 11.68 km. For hyperonic matter with $\Delta$-admixtured, the EoS predicts a maximum mass of 2.00\,$M_{\odot}$ and a radius of 11.08 km. All these MR relations satisfy the mass constraints from PSR J0740-220 and several radius constraints from NICER measurements~\cite{miller2021,2021ApJ...918L..27R, Miller_2019a, Riley_2019}, including the recent one for PSR J0437-4715~\cite{Choudhury:2024xbk}. The solid dot represents the last stable point reached in the center of the maximum-mass solution of the TOV equation. The dashed line after the solid dot corresponds to the unstable part. 

The right plot shows the EoS with a phase transition. The solid lines correspond to the hadronic matter followed by a branch of hybrid stars, represented by dashed lines. The star symbol marks the hadron-quark phase transition point. The solid dot represents the last stable point reached in the center of the maximum-mass solution of the TOV equation. The inset shows a zoomed plot version at around the maximum mass. For the hybrid EoS with nucleons only, the maximum mass is 2.29\,$M_{\odot}$ with a radius of 13.02 km. Since the phase transition to the quark matter occurs at high density, a small part of the MR relation presents hybrid stars before it reaches the unstable branch. Including delta baryons soften the EoS and hence the maximum mass decreases to 2.25\,$M_{\odot}$ only and the radius to 11.81 km, thereby representing a very small hybrid stars branch. The radius at the canonical mass, $R_{1.4}$ is 13.47 km for nucleons and 12.97 km for nucleons with delta baryons. So while the maximum decreases by around 0.17\,$M_{\odot}$ for nucleonic only EoS when phase transition is considered, this decrease is very small for N+$\Delta$ EoS, $\approx$ 0.05\,$M_{\odot}$. This is because deltas appear at a very high density and the phase transition takes place at a much higher density, allowing for a very small amount of quark matter in the core compared to the pure nucleonic hybrid EoS. 

For the hybrid EoS with nucleons and hyperons, the maximum mass is 1.95\,$M_{\odot}$ with a radius of 12.54 km. We have a substantial amount of pure quark phase here, as the phase transition point is at low density in comparison to all other EoSs. Adding delta baryons slightly increases the maximum mass to 1.98\,$M_{\odot}$ because of the delayed phase transition, with a smaller radius of 11.63 km. The MR profiles satisfy the 2.0\,$M_{\odot}$ threshold and other constraints. The hybrid nuclear EoS with and without deltas, N and N+H+$\Delta$, satisfy the 2.0\,$M_{\odot}$ limit of PSR J0740+6620. Despite selecting quark parameters for a stiff EoS, including hyperons and a phase transition to quark matter leads to an EoS that softens enough to limit the star's maximum mass to slightly under 2\,$M_{\odot}$, but satisfies the 1$\sigma$ constraint from PSR J0740+6620. 
\begin{table*}[!ht]
\centering
\caption{Stellar properties for different compositions of EoS: maximum mass ($M_{\text{max}}$) in $M_{\odot}$, radius (in km) at maximum mass ($R_{\text{max}}$), at 2.0\,$M_{\odot}$ ($R_{2.0}$), and at 1.4\,$M_{\odot}$ ($R_{1.4}$). Dimensionless tidal deformability at 1.4\,$M_{\odot}$ ($\Lambda_{1.4}$), and speed of sound squared at maximum mass configuration ($c_{s, max}^{2*}$). The upper four rows correspond to the EoS without a phase transition, while the lower rows with a phase transition. }
\begin{tabular}{cp{2.0cm}p{2.0cm}p{2.0cm}p{2.0cm}p{2.0cm}p{2.0cm}}
\hline
Composition & $M_{\text{max}}$ ($M_{\odot}$) & $R_{\text{max}}$ (km) & $R_{2.0}$ (km) & $R_{1.4}$ (km) & $\Lambda_{1.4}$ & $c_{s, max}^{2*}$  \\
\hline
\hline
N & 2.46 & 12.04 & 13.28 & 13.28 
  & 712.75 & 0.75  \\
\hline
N+$\Delta$ & 2.28 & 11.30 & 12.40 & 12.81 
  & 522.47 & 0.71  \\
\hline
N+H & 2.04 & 11.68 & 12.52 & 13.28 
  & 712.75 & 0.54  \\
\hline
N+H+$\Delta$ & 2.00 & 11.08 & 11.37 & 12.80 
  & 515.25 & 0.51  \\
\hline
\hline
N (0.90,1.25) & 2.29 & 13.02 & 13.38 & 13.47 & 712.97 & 0.27  \\
\hline
N+$\Delta$ (0.90,1.25) & 2.25 & 11.81 & 12.48 & 12.97 & 522.65 & 0.27  \\
\hline
N+H (0.65,133) & 1.95 & 12.54 & - & 13.47 & 712.97 & 0.25  \\
\hline
N+H+$\Delta$ (0.65,133) & 1.98 & 11.63 & - & 12.97 & 515.44 & 0.25  \\
\hline
\end{tabular}
\label{tab:stellar}
\end{table*}
\begin{figure}
\centering
\includegraphics[width=\linewidth]{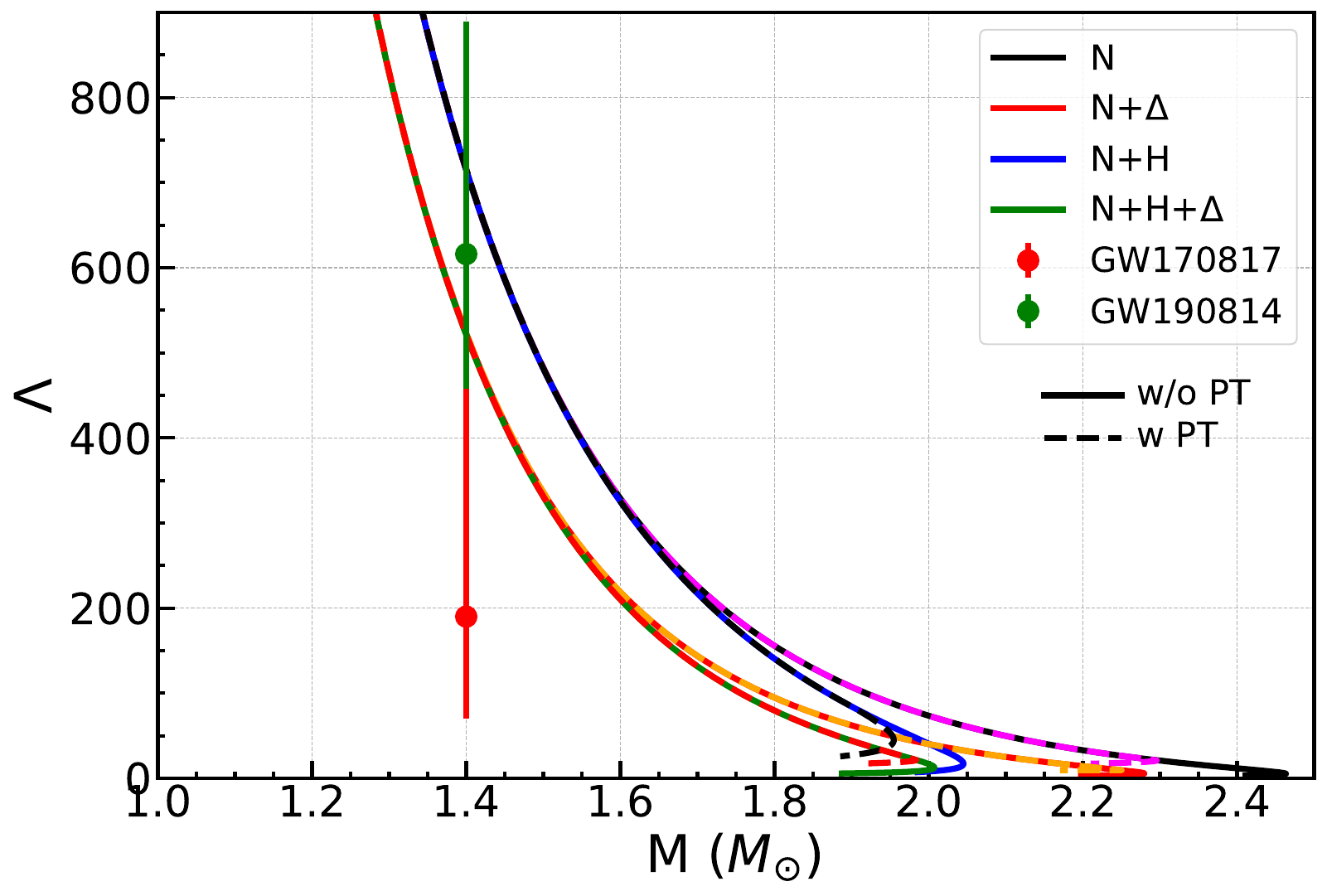}
\caption{Dimensionless tidal deformability as a function of $M$ for the EoS with different hadronic compositions. Solid (dashed) lines correspond to the EoS without (with) a phase transition to the quark matter. The red line represents the constraint on dimensionless tidal deformability at 1.4\,$M_{\odot}$ from GW170817 measurement, $\Lambda$ = 190$^{+390}_{-120}$~\cite{PhysRevLett.121.161101}, while as the green line represents the constraint on dimensionless tidal deformability at 1.4\,$M_{\odot}$ from GW190814 measurement, $\Lambda$ = 616$^{+273}_{-158}$ if the secondary component is an NS~\cite{LIGOScientific:2020zkf}. }
\label{figLambda} 
\end{figure}

Figure \ref{figLambda} shows the dimensionless tidal deformability as a function of mass for the different compositions of the EoS studied without (solid) and with (dashed) phase transition. The red and green lines represent the constraints on the dimensionless tidal deformability at 1.4\,$M_{\odot}$ from GW measurements GW170817 and GW190814, respectively $\Lambda$ = 190$^{+390}_{-120}$~\cite{PhysRevLett.121.161101} and $\Lambda$ = 616$^{+273}_{-158}$, if the secondary component is an NS~\cite{LIGOScientific:2020zkf}. 
For both the nucleon-only EoS and the nucleon-hyperon EoS, the MR relation remains unchanged at 1.4\,$M_\odot$, and the EoS including hyperons and delta resonances (N+$\Delta$ and N+H+$\Delta$) behave similarly. These characteristics are also observed when considering the hadron-quark phase transition in these EoSs.
The similarity between the curves is attributed to the density-sensitive appearance of hyperons, deltas, and/or deconfinement transition, which occur only in the densest regions near the star's core. Since the core represents a relatively small portion of the star's total volume, and tidal deformability is primarily influenced by the outer layers of the object, these exotic compositions have little effect on the star’s response to external tidal forces (see~\cite{PhysRevC.95.015801} and references therein for further discussion).
Hence the dimensionless tidal deformability goes to around 712 for N and N+H EoS with and without phase transition, satisfying the limit from GW190814. For other EoS, N+$\Delta$ and N+H+$\Delta$,  this value decreases to around 520 which is well below the limit from GW170817. All the stellar properties for the EoS without and with phase transition are presented in Table \ref{tab:stellar}.

\subsection{$f$-mode frequency: GR framework}
\label{frequency}

Since the primary goal of this work is to study non-radial $f$-mode oscillations for various hadronic EoSs—with and without a phase transition to quark matter—using a full GR treatment, we focus on presenting results obtained within the GR framework here. Results using the Cowling approximation are provided in the Appendix for reference. Where relevant, comparisons with Cowling results are discussed in the main text.
 
\begin{figure}[!t]
\includegraphics[width=\linewidth]{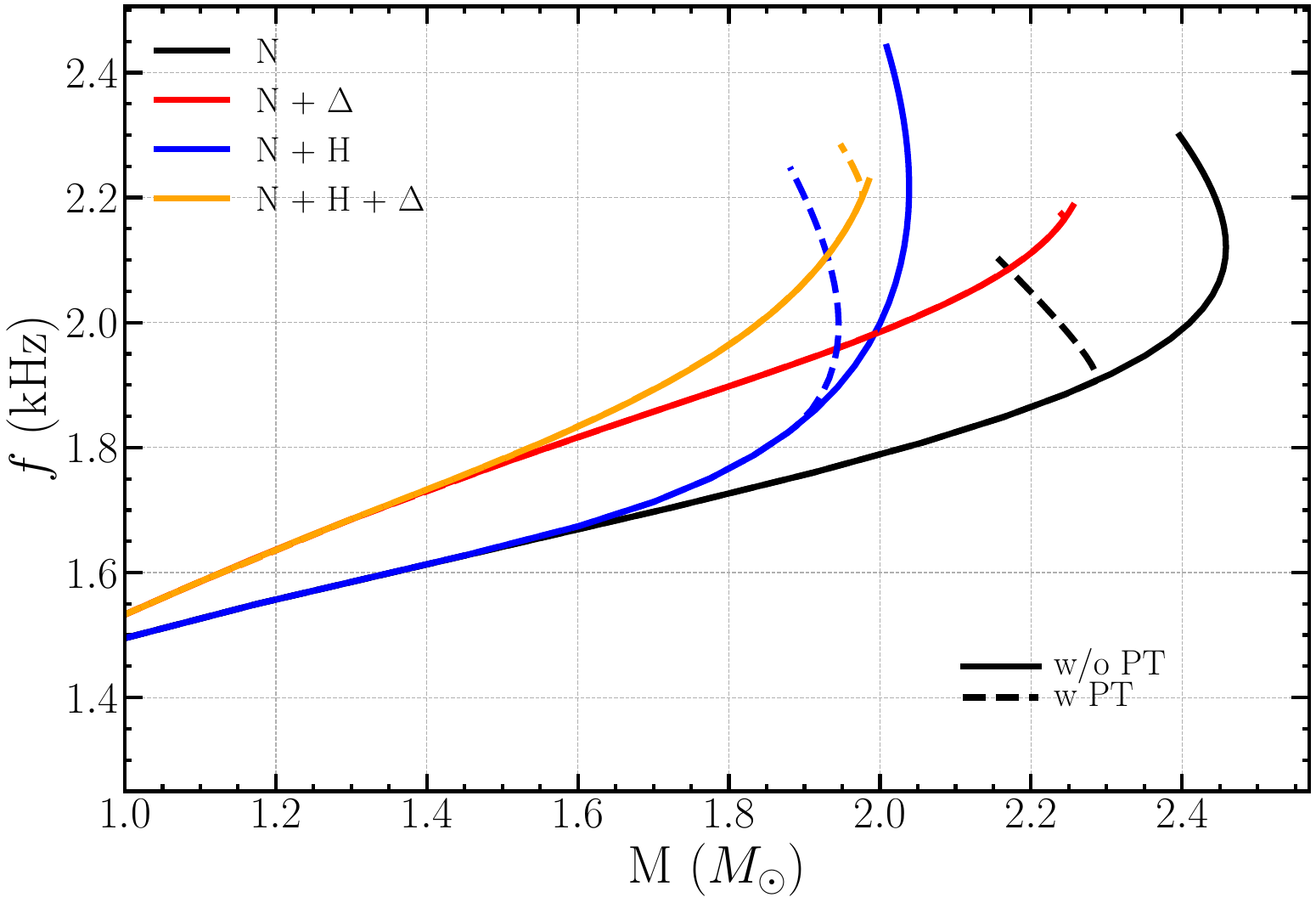}
\caption{Mass vs fundamental frequency of non-radial oscillation modes for EoS with different hadronic compositions with full GR treatment. Solid (dashed) lines represent results without (with) phase transition to the quark matter for different quark model parameters ($C, D^{1/2}$) as discussed in the text.  }
		\label{frequency_mass}
        \end{figure}

\begin{table}
\centering
\caption{Comparison between the fundamental frequencies (in kHz) calculated using GR ($f_\text{GR}$) and Cowling approximation ($f_\text{cow}$), at 1.40\,$M_{\odot}$, 1.80\,$M_{\odot}$, and at the maximum mass without and with phase transition. The percentage error (P. E.) between GR and Cowling is also shown.}
\begin{tabular}{cp{0.06\textwidth}p{0.07\textwidth}p{0.07\textwidth}p{0.07\textwidth}}
\hline
Composition & Mass & $f_\text{GR}$ & $f_\text{cow}$ & P. E. (\%) \\
\hline
\hline
 & 1.40 & 1.6130 & 2.0491 & 27.03 \\
N  & 1.80 & 1.7266 & 2.1317 & 23.46 \\
  & 2.46 & 2.1205 & 2.3734 & 11.93 \\
\hline
 & 1.40 & 1.7312 & 2.1520 & 24.31 \\
N+$\Delta$  & 1.80 & 1.8980 & 2.2908 & 20.70 \\
  & 2.28 & 2.1865 & 2.4915 & 13.95 \\
\hline
 & 1.40 & 1.6133 & 2.0491 & 27.01 \\
N+H  & 1.80 & 1.7666 & 2.1593 & 22.23 \\
  & 2.04 & 2.2043 & 2.4811 & 12.56 \\
\hline
 & 1.40 & 1.7328 & 2.1529 & 24.24 \\
N+H+$\Delta$  & 1.80 & 1.9644 & 2.3419 & 19.22 \\
  & 2.00 & 2.2269 & 2.5422 & 14.16 \\
\hline
\hline
  & 1.40 & 1.6130 & 2.0491 & 27.03 \\
N (0.90,1.25)  & 1.80 & 1.7266 & 2.1317 & 23.46 \\
  & 2.29 & 1.9110 & 2.2441 & 17.43 \\
\hline
  & 1.40 & 1.7300 & 2.1520 & 24.39 \\
N+$\Delta$ (0.90,1.25)  & 1.80 & 1.8980 & 2.2908 & 20.70 \\
  & 2.25 & 2.1662 & 2.4876 & 14.84 \\
\hline
 & 1.40 & 1.6133 & 2.0491 & 27.01 \\
N+H (0.65,133)  & 1.80 & 1.7666 & 2.1599 & 22.27 \\
  & 1.95 & 2.0018 & 2.3166 & 15.73 \\
\hline
 & 1.40 & 1.7329 & 2.1529 & 24.24 \\
N+H+$\Delta$ (0.65,133)  & 1.80 & 1.9648 & 2.3422 & 19.21 \\
  & 1.98 & 2.2018 & 2.5225 & 14.57 \\
\hline
\end{tabular}
\label{tab:frequency_comparison}
\end{table}

Figure \ref{frequency_mass} illustrates the relationship between $f$-mode frequencies and neutron star mass for various stellar compositions within the full GR treatment. The solid lines represent results without phase transition (w/o PT), while the dashed lines represent the ones with phase transition to the quark matter (w PT) for different quark model parameters ($C, D^{1/2}$) as discussed earlier. For the pure nucleonic EoS, the $f$-mode frequency at maximum mass configuration (2.45\,$M_{\odot}$) reaches 2.12 kHz within the GR framework, which decreases to a value of 1.61 kHz at the canonical mass of 1.4\,$M_{\odot}$. The inclusion of exotic particles (hyperons and $\Delta$ baryons) systematically affects these frequencies, with each additional exotic component softening the EoS in different ways. This softening reduces the maximum mass while increasing the corresponding $f$-mode frequencies in all cases, both with and without phase transitions. The phase transition to quark matter creates a distinct signature in the mass-frequency relationship, especially at higher masses, due to the maximum masses consistently occurring in the hybrid star branch.
 These results align with previous findings in~\cite{Pradhan:2022vdf,Kunjipurayil:2022zah,Roy:2023gzi}, and the complete comparative analysis between the Cowling approximations and GR formalism across all compositions is summarized in Table~\ref{tab:frequency_comparison}. 

\begin{figure}[!t]
\includegraphics[width=\linewidth]{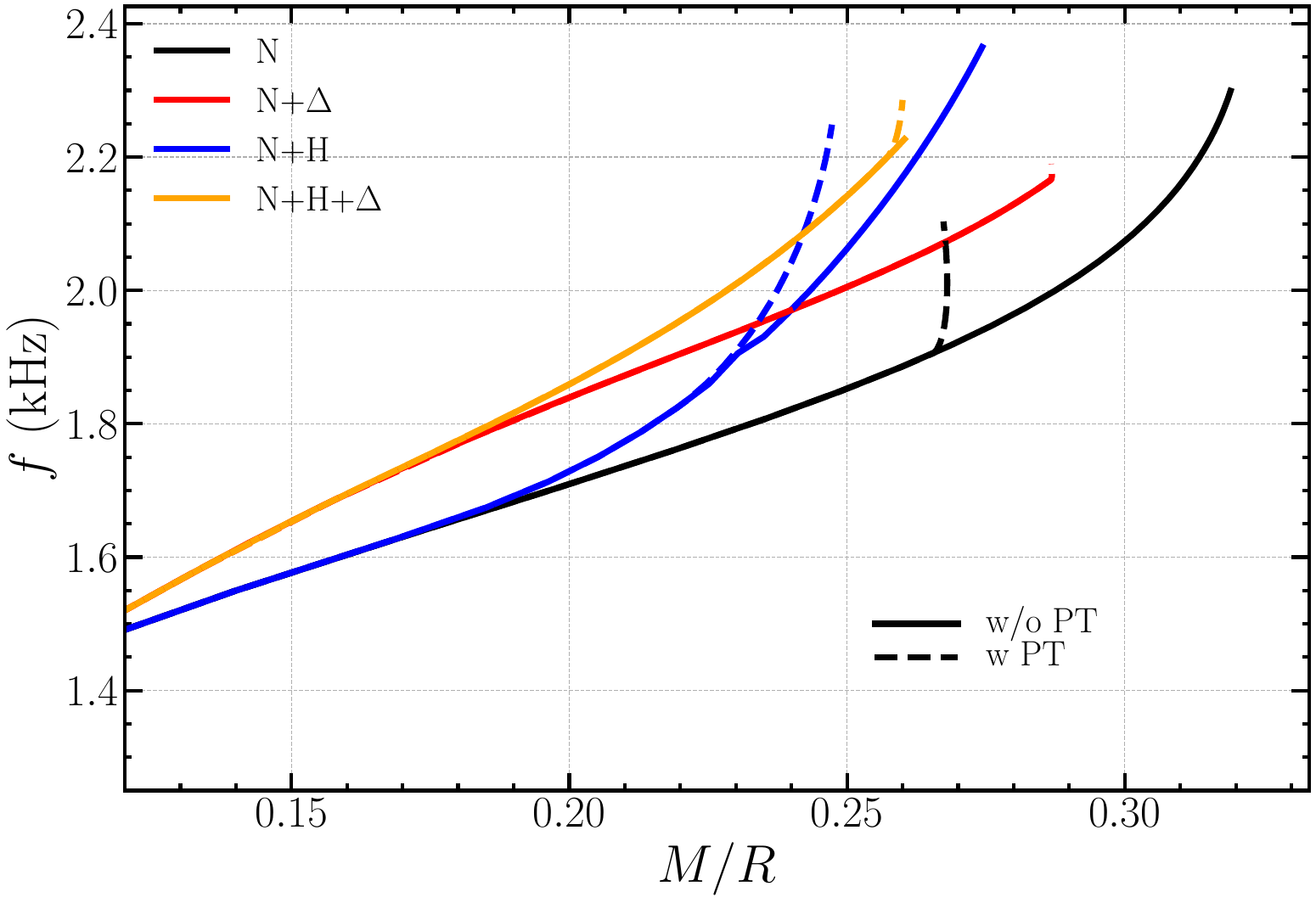}
\caption{Stellar compactness ($C = M/R$) vs fundamental frequency of non-radial oscillation modes for EoS with different hadronic compositions. Solid (dashed) lines represent results without (with) phase transition to the quark matter.  }
		\label{figCf}	 	
     \end{figure}

From Figure~\ref{frequency_mass}, we see that at $1.4\, M_\odot$, the $f$-mode frequencies are nearly identical across all compositions, indicating that exotic degrees of freedom have little impact at lower masses. However, at $1.8\, M_\odot$, a clear hierarchical pattern emerges, as shown in Table~\ref{tab:frequency_comparison}, with the ordering $f_{\text{N}} < f_{\text{N}+\text{H}} < f_{\text{N}+\Delta} < f_{\text{N}+\text{H}+\Delta}$. This trend suggests that the inclusion of hyperons and $\Delta$ baryons systematically increases the $f$-mode frequency, reflecting their growing influence on the neutron star’s oscillatory behavior. However, as we approach the maximum mass, this ordering does not strictly hold. In particular, the $f$-mode frequency for the N+H case slightly exceeds that of N+$\Delta$ case by a very small margin ($\approx$ 0.02 kHz), indicating a minor reversal in the earlier hierarchy. Interestingly, in the presence of a phase transition to the quark matter (dashed lines), the hierarchical trend remains consistent even near the maximum mass. This is because the transition to quark matter occurs at a lower central density in the N+H case than in the N+$\Delta$ case, leading to a more significant change in mass. As a result, the frequency for N+H drops more compared to N+$\Delta$, preserving the overall ordering. 

This systematic variation in the $f$-mode frequency mainly arises from changes in the star’s compactness and internal density profile. Since $f$-mode oscillations are characterized by surface-dominated fluid perturbations and core-dominated metric perturbations~\cite{Kunjipurayil:2022zah}, compactness plays a key role in determining the oscillation properties. Stiffer EoSs, associated with larger radii and lower mean densities, result in weaker restoring forces and lower $f$-mode frequencies. Conversely, softer EoSs lead to smaller radii, steeper density gradients, and higher mean densities, all contributing to higher $f$-mode frequencies. The presence of hyperons and $\Delta$ baryons soften the EoS, reducing pressure support, which strengthens the restoring force for fluid perturbations. It also affects the effective sound speed and density stratification, further influencing the oscillation dynamics. The increased compactness in these cases enhances gravitational coupling and reduces damping times, facilitating stronger gravitational wave emission~\cite{Burgio:2011qe,Kunjipurayil:2022zah,Lindblom:1983ps}.

Using GW frequencies to distinguish different EoS families can be effective by considering variations with star compactness, which can be independently assessed through gravitational redshift measurements from spectral line observations~\cite{Andersson:1996pn, Andersson:1997rn, Benhar:2004xg, Glendenning:1997wn}.

Figure~\ref{figCf} illustrates the variation of $f$-mode frequencies with compactness ($C$), i.e., the $f$–$C$ relation, for EoS with different compositions. The solid lines represent results without phase transitions, while the dashed lines represent results with a phase transition to quark matter.

While compactness determines the overall gravitational binding, exotic degrees of freedom alter the internal structure, modifying the restoring forces for oscillations. Purely nucleonic stars exhibit the lowest $f$-mode frequencies due to their shallower density profiles, whereas those containing hyperons and $\Delta$ baryons become more centrally condensed. Even at fixed compactness, composition plays a crucial role in shaping oscillation properties, leading to systematically higher $f$-mode frequencies in EoSs with exotic matter. 
The presence of a phase transition alters the $f$-$C$ relationship. For a given compactness value, the frequencies in the phase transition models tend to be different from their non-phase transition counterparts. This reflects the fundamental changes in the EoS when quark matter appears in the stellar core. The phase transition models show a more limited range of stable compactness values as compared to the one without. This is consistent with the understanding that phase transitions generally soften the EoS, reducing the maximum stable mass and altering the mass-radius relationship. Our results align with those of Ref.~\cite{Kalita:2023rbz}, but our study explicitly accounts for full general relativistic effects, ensuring a more accurate description of NS oscillations.

 \begin{figure}[t]	 		
\includegraphics[width=\linewidth]{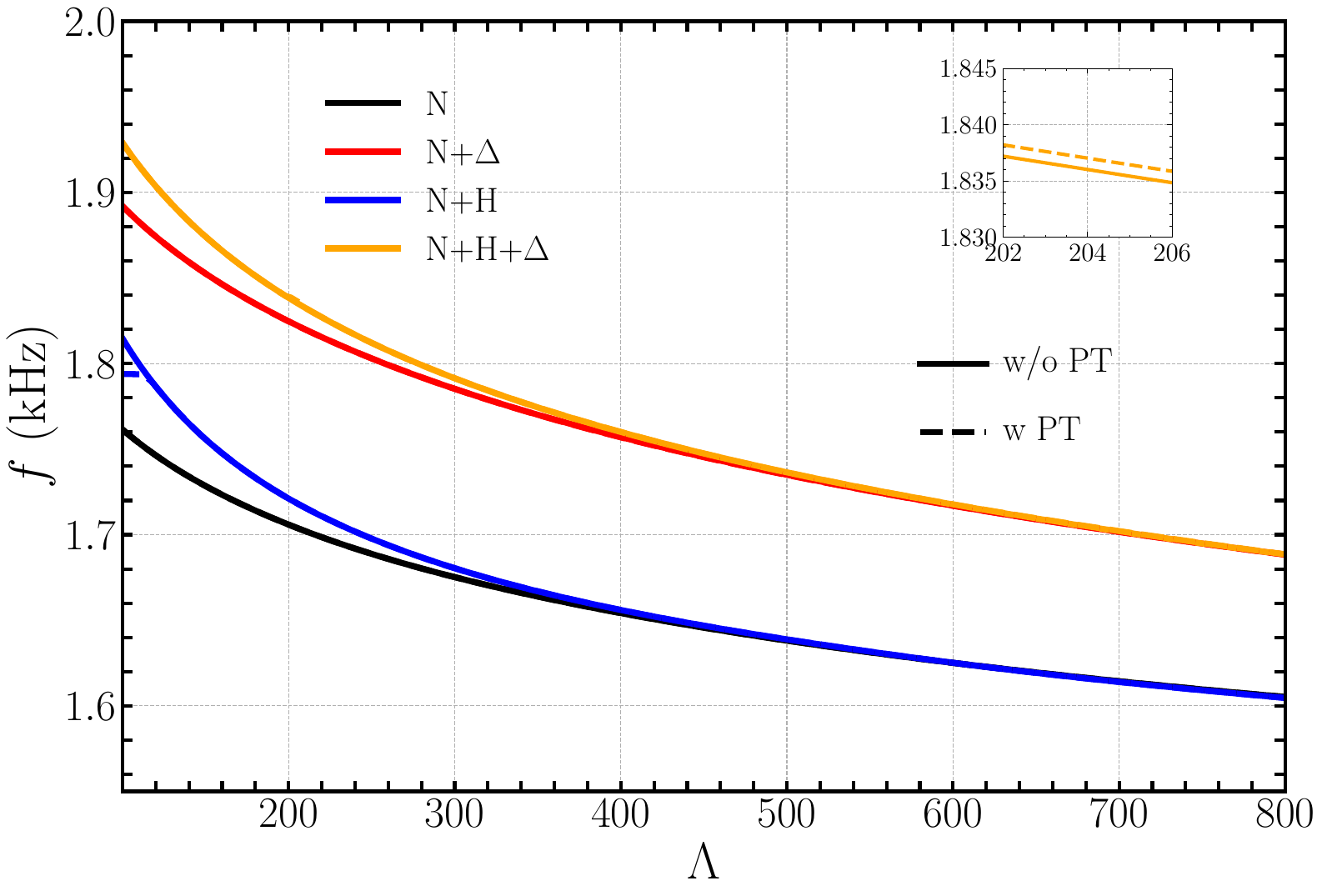}
\caption{Dimensionless tidal deformability ($\Lambda$) vs fundamental frequency of non-radial oscillation modes for EoS with different hadronic compositions. Solid (dashed) lines represent results without (with) phase transition to the quark matter. The inset shows the low $\Lambda$ region in detail. }
		\label{figLambdaf}	 	
     \end{figure}
     
Besides the basic properties such as mass, radius, and compactness, dimensionless tidal deformability serves as a crucial observable for constraining the NS EoS.
By separately measuring the $f$-mode frequency and tidal deformability, we obtain insightful data that enhances our understanding of NSs internal structure.
 Figure~\ref{figLambdaf} illustrates the relationship between $f$-mode frequencies and the dimensionless tidal deformability. Our results for the $f$-mode frequency lie well within the limits obtained from the GW170817 observation which is estimated between 1.43 kHz and 2.90 kHz for the more massive NS and between 1.48 kHz and 3.18 kHz for the less massive one. Furthermore, analysis of the $f$-mode frequencies with respect to tidal deformability reveals a convergence phenomenon: beyond a tidal deformability parameter $\Lambda \approx 300$, the $f$-mode frequencies become effectively indistinguishable across all studied compositions, as illustrated in Figure~\ref{figLambdaf}. This convergence persists regardless of whether phase transitions to quark matter are present in the EoS. Such behavior indicates that in this high-deformability regime, the $f$-mode oscillations no longer serve as effective discriminators between EoSs with and without phase transitions, suggesting that the influence of compositional differences on oscillation properties becomes negligible at these deformability values. The inset shows a more detailed description of the plot at low tidal deformability, to distinguish between the results without and with phase transition to the quark matter.

\section{Gravitational Wave Asteroseismology-Universal relations}
\label{GW}


Neutron star asteroseismology aims to connect the oscillation modes' angular frequencies and GW damping timescales to the star’s core properties, including mass, radius, and rotational frequency. By using inverse asteroseismology, it is possible to derive relationships that are largely independent of the specific EoS. This approach leverages GW observations in combination with the star’s global properties--particularly its rotational frequency, which plays a crucial role in rapidly rotating neutron stars--to infer internal structure and dynamics.
The concept of GW asteroseismology was initially introduced by~\citet{Andersson:1996pn} for certain polytropic EoSs and later explored for some realistic EoSs~\cite{Andersson:1997rn}. They derived an empirical asteroseismology relation between $f$-mode frequency as a function of average density of the star, namely,
\begin{equation}
 f (\text{kHz}) = a+b \sqrt{ \frac{\bar{M}}{{\bar{R}^3}} }, 
 \label{eq:fit}
\end{equation}
in terms of the dimensionless parameters $\bar{M} = \frac{M}{1.4\,M_{\odot}}$ and $\bar{R}=\frac{R}{10\,\text{km}}$. This was further probed with some EoSs containing exotic phases such as hyperons and quarks by~\citet{Benhar:2004xg}. More studies with exotic phases, quarks, and dark matter were also carried out in Ref.~\cite{Blazquez-Salcedo:2013jka, Ranea-Sandoval:2018bgu, Shirke:2024ymc, PhysRevD.104.123006, Pradhan:2020amo}.
But no work in the context of $\Delta$ baryons has been carried out for the $f$-mode frequency, especially with a hadron-quark phase transition considered. 
To facilitate a comprehensive comparison with prior studies and provide detailed discussion, we include Cowling approximation results alongside the GR results, as this approach has been utilized in the literature.

\begin{figure}[!t]		 		
  \includegraphics[width=\linewidth]{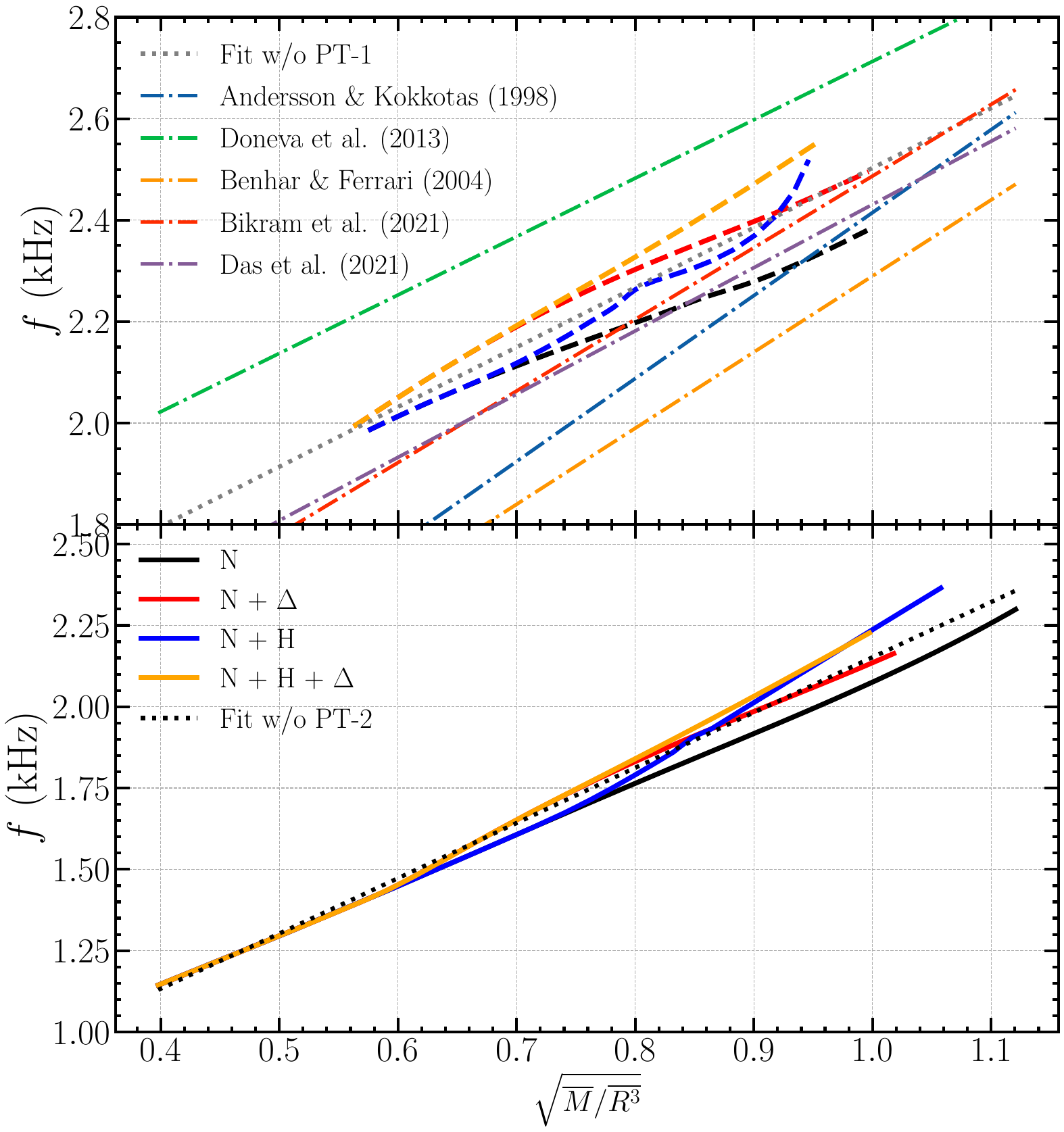}
\caption{Average density of the star vs fundamental frequency of non-radial oscillation modes for EoS with different hadronic compositions. The lower (upper) plot represents results from the full General Relativistic (Cowling) treatment. The dot-dashed lines in the upper plot correspond to the fits from various studies whereas the dotted line corresponds to the fit from our work.}
		\label{figMRf1}	 	
     \end{figure}

\begin{figure}[!t]
\includegraphics[width=\linewidth]{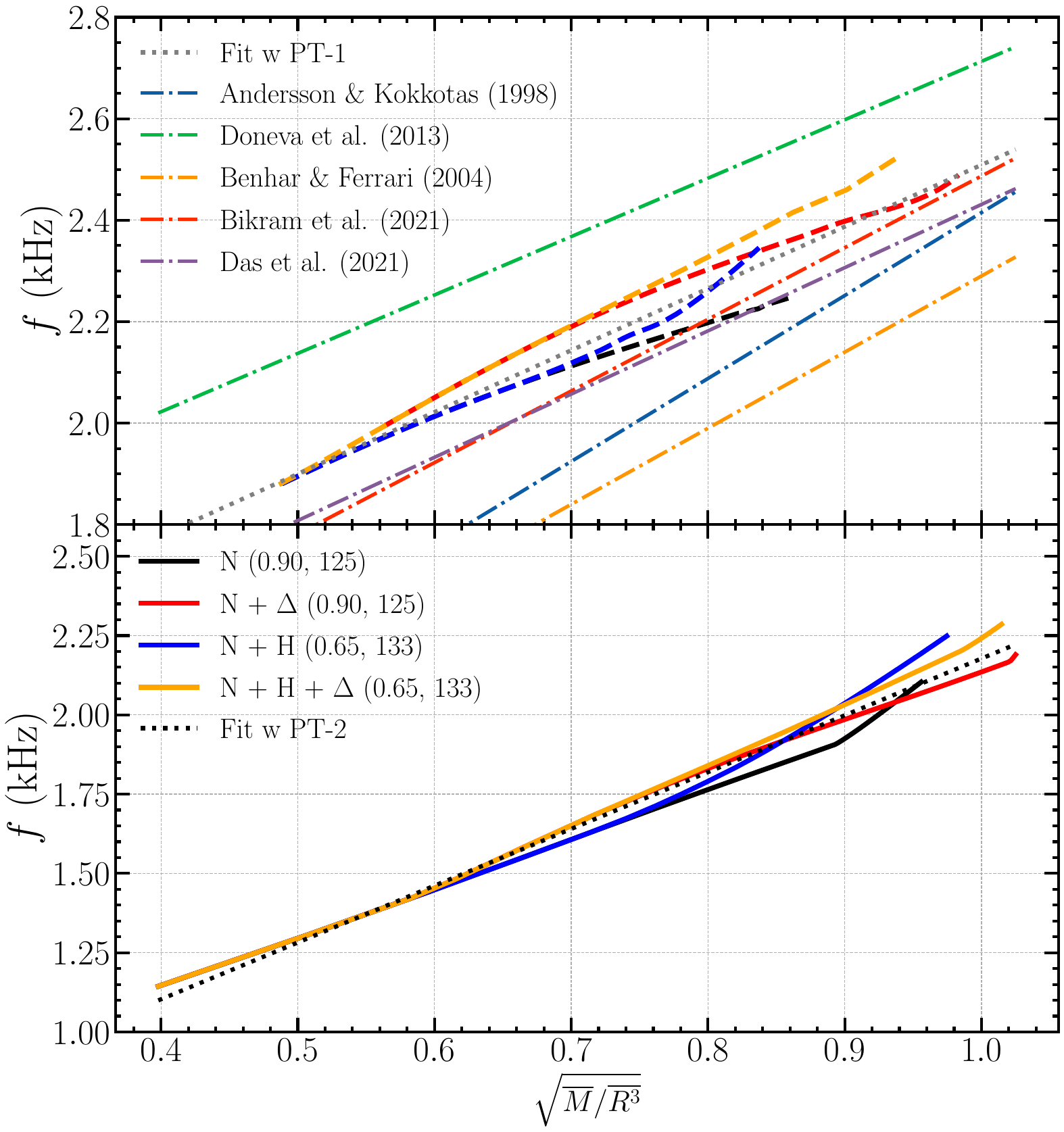}
\caption{Same as Figure \ref{figMRf1}, but with phase transition to the quark matter at different quark model parameters ($C, D^{1/2}$).}
		\label{figMRf2}	 	
     \end{figure}

In Figures \ref{figMRf1} and \ref{figMRf2}, we present the empirical asteroseismology relation for $f$-mode frequencies as a function of the average density of the star, respectively for the scenarios without and with a phase transition. The upper panels present fitting relations based on the Cowling approximation, while the lower panels display results from the full GR framework. Dot-dashed lines represent fits from previous studies~\cite{Andersson:1997rn, Doneva:2013zqa, Benhar:2004xg, Pradhan:2020amo, PhysRevD.104.123006}, and the dotted line corresponds to the fit obtained from our work. All the different values of $a$ and $b$ for the above-fit relation are shown in Table \ref{tab:fit}. For the Cowling approximation fit, the values of $a$ and $b$ from our fit are 1.32 and 1.18 kHz, respectively, without a phase transition, and 1.29 and 1.22 kHz, with a phase transition. They are named as Fit w/o PT-1 and Fit w PT-1 for without and with phase transition, respectively.
Unlike earlier works, our results differ significantly from previous studies because we included $\Delta$ baryons in our analysis. This consideration alters the equation of state, leading to the observed variations in the fit relations. For the GR fit, the values of $a$ and $b$ are 0.44 and 1.72 kHz, respectively, without phase transition, and 0.39 and 1.79 kHz with phase transition.
They are named Fit w/o PT-2 and Fit w PT-2 for without and with phase transition, respectively. As discussed in Ref.~\cite{Pradhan:2022vdf, PhysRevD.70.124015, PhysRevC.99.045806}, while empirical relations are designed to be independent of the underlying EoS, they still retain some degree of model dependence. Given that the NS masses are among the most precisely measured global properties, their combination with mode frequency observations can aid in distinguishing between different EoS models and provide insights into the behavior of matter at high densities. In essence, these empirical fits serve not only as tools for estimating global observables but also as a means to constrain EoS stiffness and identify possible signatures of exotic matter. \citet{Pradhan:2020amo} obtained values of $a = 1.075$ and $b = 1.412$ using the Cowling approximation for a nucleonic-hyperonic composition. In contrast, our fit yields $a = 1.32$ and $b = 1.18$ (Fit w/o PT-1), which is expected due to the additional presence of $\Delta$ baryons in our EoSs. This difference highlights the impact of $\Delta$ baryons on the fit parameters. Under full GR calculation, our fitted relation gives $a = 0.44$ and $b = 1.72$ (Fit w/o PT-2), whereas \citet{Pradhan:2022vdf} report 0.535 and 1.643.

\begin{table}[!ht]
\centering
\caption{Values of fitting coefficients $a$ and $b$ in kHz for Eq.~\eqref{eq:fit} from different works and our results.}
\begin{tabular}{lcc}
\hline
Reference & $a$ (kHz) & $b$ (kHz)  \\
\hline
\citet{Benhar:2004xg} & 0.79 & 1.500 \\
\citet{Andersson:1997rn} & 0.78 & 1.635 \\
\citet{PhysRevD.104.123006} & 1.185 & 1.246 \\ 
\citet{Pradhan:2020amo} & 1.075 & 1.412 \\
\citet{Doneva:2013zqa} & 1.562 & 1.151 \\
\hline
Our results\\
\hline
 Fit w/o PT-1  & 1.32 & 1.18   \\
 Fit w/o PT-2  & 0.44 & 1.72  \\
 Fit w PT-1  & 1.29 & 1.22    \\
 Fit w PT-2  & 0.39 & 1.79  \\
\hline
\end{tabular}
\label{tab:fit}
\end{table}


Unlike the fitting relations in Eq.~\eqref{eq:fit} which exhibit some model dependence, certain universal relations remain largely independent of the EoS. These relations are particularly useful for inferring the NS properties from QNM observations. Previous studies from Ref.~\cite{Pradhan:2020amo, Pradhan:2022vdf} have demonstrated that the mass-scaled angular frequency $\omega M$ follows a universal trend with stellar compactness.
Extending this idea, prior research on $g$-modes has shown a similar universal relation between $\omega M$ and compactness, $M/R$~\cite{Sotani_2011}. In this work, we examine how these relations are influenced by the presence of hyperons, $\Delta$ baryons, and their combination along with a phase transition to the quark phase, focusing on the behavior of angular frequency when scaled by mass and radius.

     \begin{figure}[t]	 		
\includegraphics[width=\linewidth]{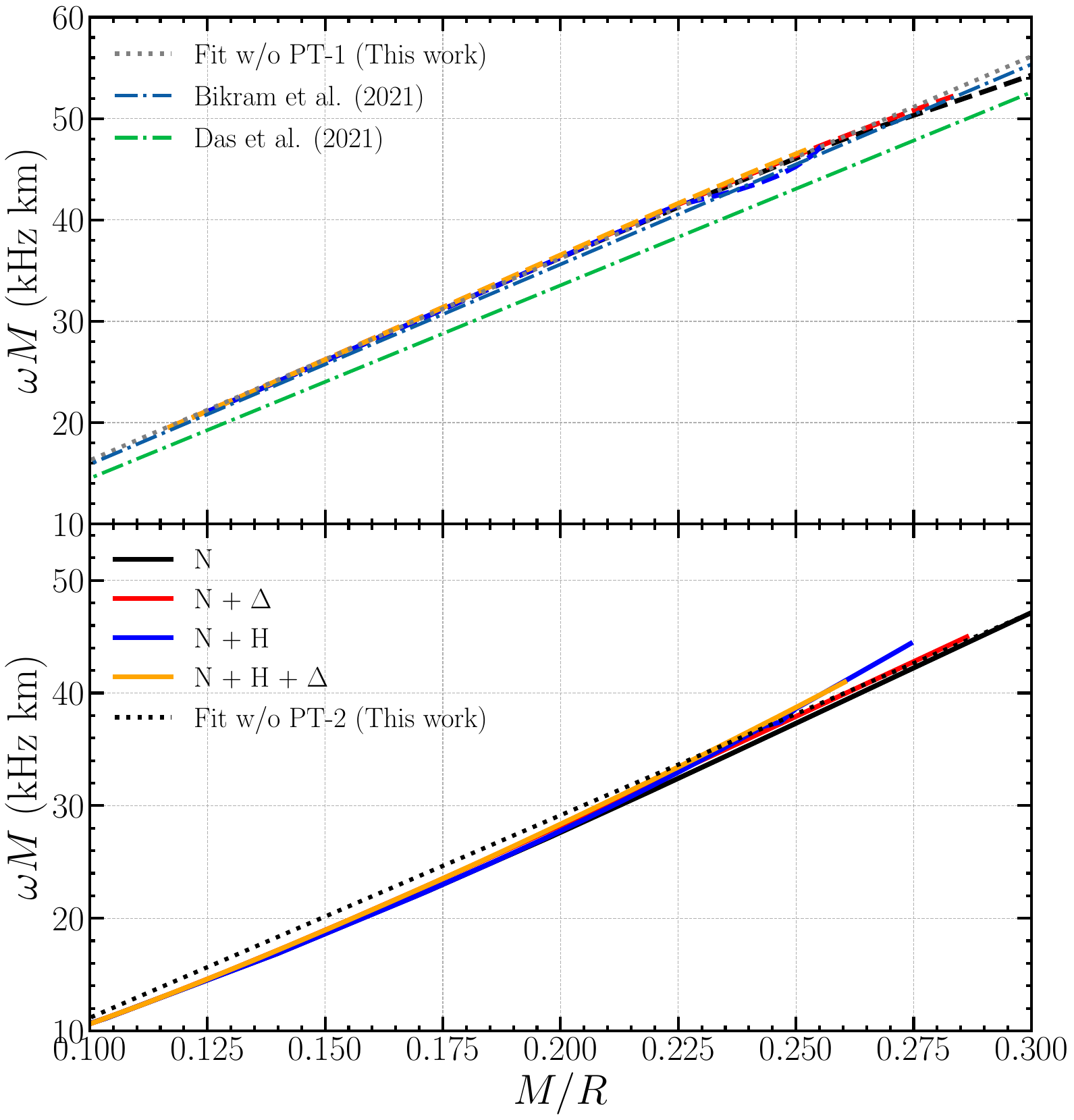}
\caption{Stellar compactness vs the angular frequency ($\omega=2\pi f$) scaled by mass ($\omega M$) for EoS with different hadronic compositions. The lower (upper) plot represents results from the full general relativistic (Cowling) treatment. The dot-dashed lines in the upper plot correspond to the fits from various studies, whereas the dotted line in both the upper and lower plot corresponds to the fit from our work.}
		\label{figComegaM1}	 	
     \end{figure}

     \begin{figure}[t]	 		
\includegraphics[width=\linewidth]{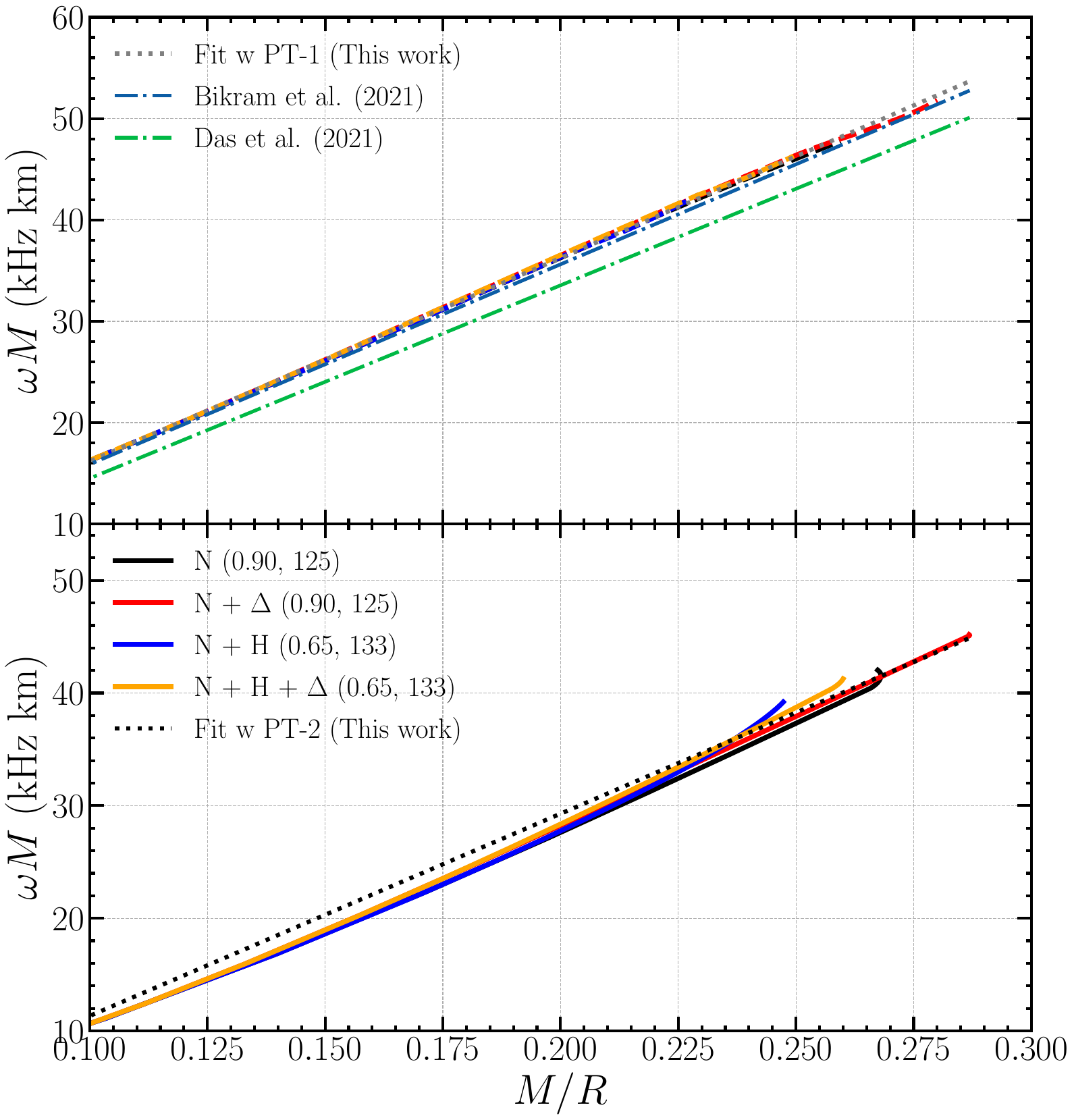}
\caption{Same as Figure \ref{figComegaM1}, but with phase transition to the quark matter at different quark model parameters ($C, D^{1/2}$). }
		\label{figComegaM2}	 	
     \end{figure}

In Figures~\ref{figComegaM1} and \ref{figComegaM2}, we present the mass-scaled angular frequency ($\omega M$) as a function of stellar compactness. Figure~\ref{figComegaM1} shows the analysis without phase transitions, whereas Figure~\ref{figComegaM2} includes a phase transition to quark matter. The lower (upper) panels display the results from the full general relativistic (Cowling) treatment. The universal relation between $\omega M$ and $M/R$ is given by
\begin{equation}
\label{equation_uni}
\omega M =a \left( \frac{M}{R} \right) - b ,
\end{equation}
where $a$ and $b$ are fitting coefficients in kHz$\cdot$km. In the upper plot, dot-dashed lines represent fits from various studies, while the dotted line in both upper and lower plots corresponds to the fit derived from our work. Although the fit from~\citet{Pradhan:2020amo} matches very closely to our fit, the one from~\citet{PhysRevD.104.123006} differs. This can be attributed to the fact that their study incorporates the presence of dark matter, whereas ours focuses on the inclusion of delta baryons in the composition. The values of $a$ and $b$ obtained in these references are provided, along with our results, in Table~\ref{tab:omegaMfits}.
\begin{table}[htb]
\centering
\caption{Values of the fitting coefficients $a$ and $b$ for Eq.~\eqref{equation_uni} from different works and from our results.}
\begin{tabular}{lcc}
\hline
Reference & $a$ (kHz$\cdot$km) & $b$ (kHz$\cdot$km) \\
\hline
\citet{PhysRevD.104.123006} & 190.48 & $2.98$ \\
\citet{Pradhan:2020amo} & 197.30 & $3.84$ \\
\hline
Our results\\
\hline
Fit w/o PT-1 & 199.40 & $3.66$ \\
Fit w PT-1   & 200.00 & $3.88$ \\
Fit w/o PT-2 & 179.61 & $6.63$ \\
Fit w PT-2   & 180.65 & $6.92$ \\
\hline
\end{tabular}
\label{tab:omegaMfits}
\end{table}
These numbers show that empirical fits developed for purely nucleonic stars or the ones with dark matter may not be accurate when $\Delta$-baryons are considered. The effect of $\Delta$-baryons in the case of a phase transition leads to a more noticeable deviation in the fit compared to the ones reported in Refs.~\cite{Pradhan:2020amo, PhysRevD.104.123006}. This implies that the fits from this work in both the Figures \ref{figComegaM1} and \ref{figComegaM2} accurately capture these changes while other fits do not. Such a finding emphasizes the need for an updated empirical relation incorporating both $\Delta$-baryons and phase transitions.

\begin{figure}[!htb]	 		
  \includegraphics[width=\linewidth]{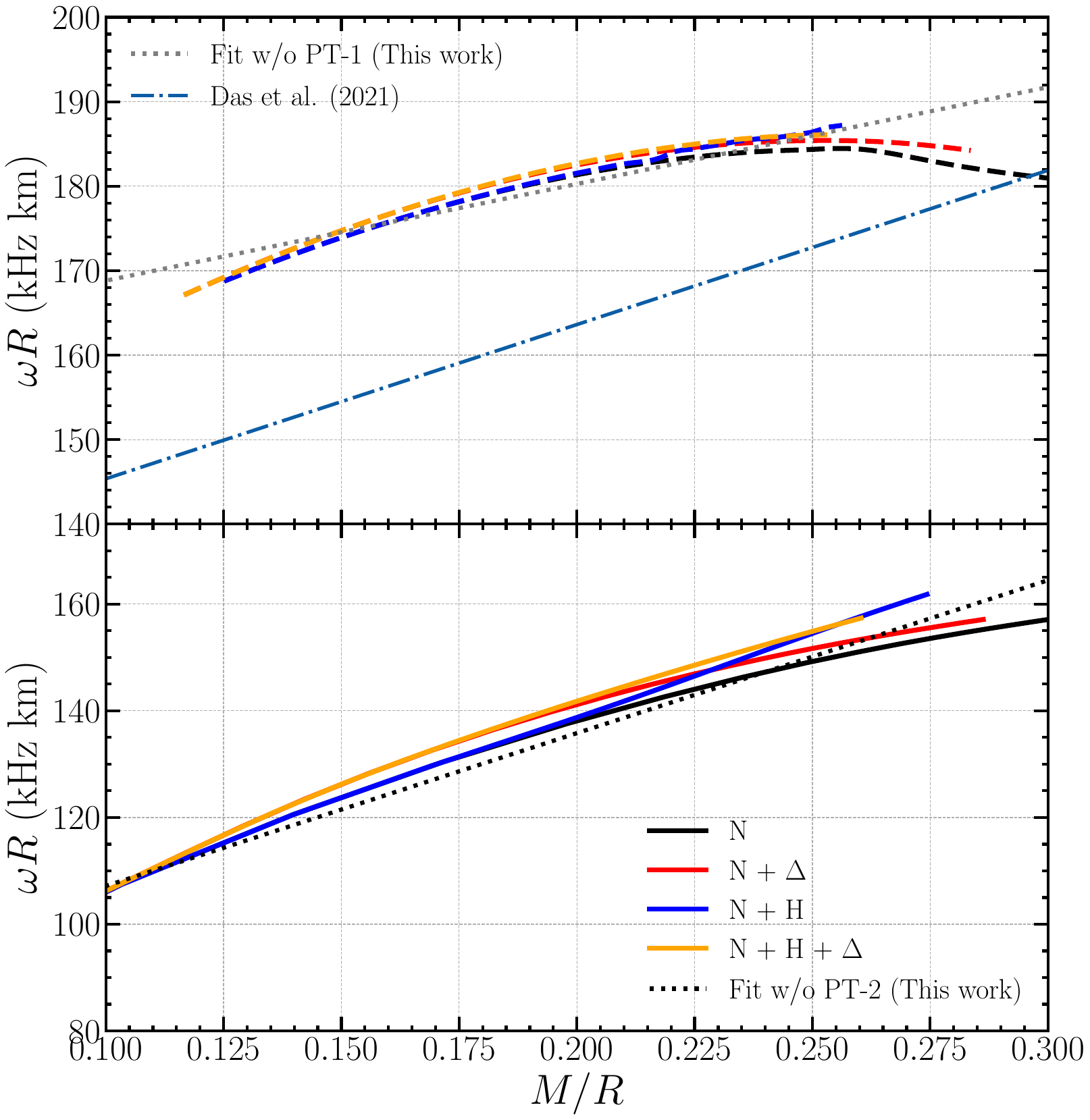}
\caption{Same as Figure \ref{figComegaM1} but angular frequency scaled by radius ($\omega R$) as function of compactness.}
		\label{figComegaR1}	 	
\end{figure}

\begin{figure}[!htb]	 		
\includegraphics[width=\linewidth]{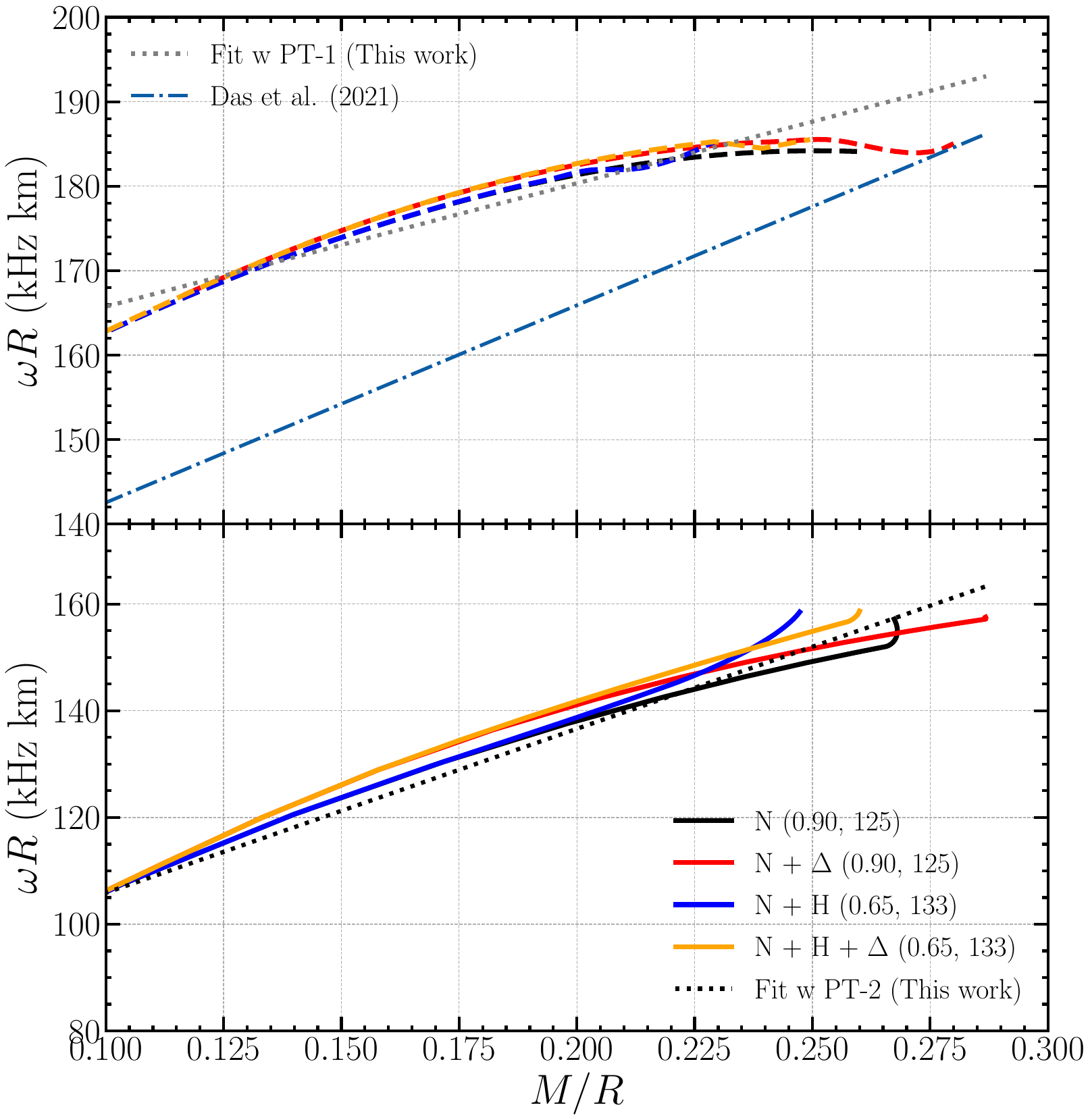}
\caption{Same as Figure \ref{figComegaM2} but angular frequency scaled by radius ($\omega R$) as function of compactness.}
		\label{figComegaR2}	 	
\end{figure}

Figures \ref{figComegaR1} and  \ref{figComegaR2} illustrate the relationship between $\omega R$ (the product of the $f$-mode frequency $\omega$ and radius $R$) and the compactness ($M/R$), respectively without and with a hadron-quark deconfinement transition. The universal relation takes the same form as Eq. \eqref{equation_uni},
\begin{equation}
\omega R = a \left( \frac{M}{R} \right)+b.
\label{omegaR}
\end{equation}
The lower panels depict results obtained from full GR calculations, which account for both fluid and gravitational perturbations, providing the most accurate theoretical predictions. In contrast, the upper panels show results under the Cowling approximation, where gravitational perturbations are neglected.  This simplification leads to an overestimation of $\omega R$, evident from the consistently higher values compared to the GR results. The overestimation is more pronounced at lower compactness and reduces as compactness increases, reflecting the stronger coupling of surface fluid perturbations to the tidal field in more compact stars.
 The dotted line in both panels corresponds to the universal fit derived from the current study.  In the upper panel, the dot-dashed line represents the fit from the previous study by~\citet{PhysRevD.104.123006}, providing a comparative reference. In Table~\ref{tab:omegaRfits} we present the values of the coefficients $a$ and $b$ obtained from our fittings.
\begin{table}[!htb]
\centering
\caption{Values of the fitting coefficients $a$ and $b$ for Eq.~\eqref{omegaR} from our results.}
\begin{tabular}{lcc}
\hline
Fitting & $a$ (kHz$\cdot$km) & $b$ (kHz$\cdot$km) \\
\hline
Fit w/o PT-1 & 114.54 & 157.36 \\
Fit w PT-1   & 145.93 & 151.15 \\
Fit w/o PT-2 & 286.57 & 78.50 \\
Fit w PT-2   & 307.54 & 75.12 \\
\hline
\end{tabular}
\label{tab:omegaRfits}
\end{table}

The presence of $\Delta$-baryons modifies the relation but within the expected smooth trends of nucleonic EoS.
Their inclusion alongside a phase transition leads to a significantly altered behavior in $\omega R$, possibly introducing more abrupt changes or shifts due to density-dependent transitions. Our fits better account for these transitions and particle degrees of freedom, and offer a more refined model for neutron star oscillations.

Our results extend the findings of \citet{Pradhan:2020amo, Pradhan:2023zmg, PhysRevC.103.035810} by systematically analyzing the impact of $\Delta$ baryons on neutron star oscillations. Unlike previous works, which focused on nucleonic and hyperonic EoSs, we incorporate $\Delta$ baryons and explore their influence on $f$-mode frequencies, compactness, and tidal deformability. Our study reveals that $\Delta$ baryons introduce noticeable deviations in universal relations, particularly in the average density of the star vs $f$-mode frequency (see Figures~\ref{figMRf1} and \ref{figMRf2}) as well as in the $f$-mode frequency vs. compactness relation (see Figure~\ref{figCf}), which were not observed in prior studies. Additionally, we compare results from the Cowling approximation and full general relativity, demonstrating that the frequency discrepancies are more pronounced in EoSs containing $\Delta$ baryons, without and with phase transition to the quark matter. These differences suggest that gravitational wave asteroseismology could provide a means to detect the presence of $\Delta$ baryons in neutron stars, a possibility not considered in previous analyses.

 \section{Summary and Conclusions}
\label{summary}
In this work, we investigated the effects of $\Delta$ baryons on the equation of state (EoS), $f$-mode oscillations, and universal relations in neutron stars. Using the density-dependent relativistic mean-field (DD-RMF) model with the DDME2 parameter set, we constructed different EoSs including nucleonic, hyperonic, and $\Delta$-admixed matter. Additionally, we considered hybrid stars with a hadron-quark phase transition, modeled via the density-dependent quark mass (DDQM) approach.

Our results show that $\Delta$ baryons soften the EoS, reducing the maximum neutron star mass while modifying the stiffness at high densities. This leads to significant changes in mass-radius relations and the speed of sound in neutron stars. When incorporating a phase transition, we observe that the hybrid EoS with $\Delta$ baryons exhibits delayed quark matter onset compared to purely nucleonic or hyperonic models, influencing the hybrid star's core composition.

We confirm the well-known discrepancy between $f$-mode frequencies computed using the Cowling approximation and full GR calculation, with the Cowling approximation overestimating the frequencies by about 10\% to 30\%. While this has been shown in previous studies, our results demonstrate that the discrepancy remains significant even in the presence of additional degrees of freedom such as hyperons, $\Delta$ baryons, and a phase transition to quark matter. We observe that the discrepancy generally decreases with increasing stellar mass; however, near the maximum mass, this trend depends on the EoS. For EoSs without a phase transition, the discrepancy reduces at the maximum mass, consistent with earlier findings. In contrast, for EoSs involving a phase transition to quark matter, the discrepancy increases by a few percent compared to the case without a phase transition. This feature emphasizes the necessity of using full GR calculation to accurately model neutron star oscillations, particularly in the presence of a phase transition.

Our study provides important implications for gravitational wave detections from neutron star oscillations. The presence of $\Delta$ baryons systematically shifts the $f$~mode frequencies and modifies the empirical relations that connect them to neutron star compactness and tidal deformability. Given that current and future gravitational wave detectors (e.g., LIGO-Virgo-KAGRA, Einstein Telescope, Cosmic Explorer) aim to constrain neutron star properties with unprecedented precision, our results suggest that $\Delta$ baryons could leave measurable imprints on observed mode frequencies. Furthermore, the inclusion of $\Delta$ baryons in hybrid stars alters the expected frequency range of post-merger oscillations, which could be relevant for interpreting signals from future multi-messenger events.

In addition, we examined $f$-mode frequencies as functions of stellar compactness and tidal deformability, establishing universal relations that extend previous results. The inclusion of $\Delta$ baryons introduces deviations in these relations, suggesting potential observational signatures in gravitational wave data. Empirical fits for $f$-mode frequencies were derived for both the Cowling and the GR frameworks, demonstrating the influence of exotic baryons on neutron star oscillations and constraints on the EoS.

In this work, the quark matter parameters $C$ and $D^{1/2}$ were selected to ensure the presence of coexistence point with the chosen hadronic EoS. As outlined in Ref.~\cite{Rather:2024hmo}, the position of the phase transition is highly sensitive to these parameters. A different choice for the pair ($C, D^{1/2}$) would alter the quark EoS, which in turn would affect key stellar properties such as the maximum mass, radius, tidal deformability, and hence the oscillation frequencies. Since our results—spanning different particle compositions and phase transitions to quark matter with varying values of $C$ and $D^{1/2}$—consistently show that the universal relations, particularly those derived within the GR framework, are robust, we conclude that reasonable variations in quark matter parameters do not significantly affect these relations. The fits derived remain consistent across models, indicating that while specific stellar quantities may shift with different parameter choices, the overall universality and qualitative features of the relations will remain preserved.

In summary, our study underscores the role of $\Delta$ baryons in neutron star structure and dynamics. Their impact on $f$-mode oscillations, particularly in hybrid stars, provides new insights into dense matter physics and gravitational wave asteroseismology. These findings contribute to ongoing efforts to connect theoretical models with astrophysical observations, advancing our understanding of neutron star interiors and the QCD phase diagram.

\section{Acknowledgement}
We thank the anonymous referee for their valuable comments and suggestions, which have helped improve the quality of this manuscript.
P.T. sincerely thanks Tuhin Malik for his invaluable support in enhancing the accuracy and robustness of the code utilized in this work. This work is a part of the project INCT-FNA proc. No. 464898/2014-5. It is also supported by Conselho Nacional de Desenvolvimento Cient\'ifico e Tecnol\'ogico (CNPq) under Grants No. 307255/2023-9 (O.~L.), and 401565/2023-8 (Universal, O.~L.). I.A.R. acknowledges support from the Alexander von Humboldt Foundation. K.D.M. thanks financial support from the São Paulo State Research Foundation (FAPESP), under Grant No. 2024/01623-6.  \\

\section{appendix}
\label{appendix}

Here we describe the Cowling approximation and discuss the $f$-mode frequency calculations with and without a phase transition. The corresponding values of the frequency are compared with the GR framework and are shown in Table~\ref{tab:frequency_comparison}.

\subsection{Relativistic Cowling approximation}

In the Newtonian framework of stellar pulsations, when the perturbation of the gravitational potential caused by matter fluctuations is ignored, the resulting simplification is referred to as the Cowling approximation~\cite{1983ApJ...268..837M}. This significantly reduces the complexity of the fluid perturbation equations. Analogously, in the context of general relativity, neglecting the perturbations of the spacetime metric leads to what is known as the relativistic Cowling approximation. The relativistic Cowling equations are obtained by setting $H_0=H_1=K=0$ in Eq. (\ref{eq:ODE_DL1}), Eq. (\ref{eq:ODE_DL2}) and Eq. (\ref{eq:ODE_DL3}), and furthermore, omitting the term $-4\pi(\varepsilon+p)^2e^{(\nu+\lambda)/2}W$ in Eq. (\ref{eq:boundary_conditions_1}), leading to 
\begin{eqnarray}
\label{eq:ODE_DL3_cowling_UW}
\frac{dW}{d\ln r} & =& -(l+1)\left[W-l e^{\nu +\lambda/2} U\right] \nonumber \\
&& -\frac{e^{\lambda/2}(\omega r)^2}{c_{ad}^2}\left[U- \frac{e^{\lambda/2}\textrm{Q}}{(\omega r)^2} W\right ]\,,\\
\frac{dU}{d\ln r} && = e^{\lambda/2-\nu}\left[W -le^{\nu-\lambda/2}U\right] \,,\label{eq:ODE_DL4_cowling_UW}
\end{eqnarray}
where $W=e^{\lambda/2} r^{1-l} \xi^r$ and $U=-e^{-\nu}V=r^{-l} \omega^{-2} \delta p/(\varepsilon+p)$, $\xi^r$ are radial Lagrangian displacements defined in Eq. (\ref{eq:xi_radial}) and $\delta P$ is the Eulerian perturbation of pressure, which is related to the the Lagrangian perturbation by $\Delta P=\delta P-(\varepsilon+p) \frac{d\Phi}{dr} \xi^r$. The boundary conditions can be written explicitly as,
\begin{eqnarray}
\left.\frac{W}{U}\right|_{r=0}&=&l e^{\nu|_{r=0}} \\
\left.\frac{W}{U}\right|_{p=0}&=&\frac{\omega^2R^3}{GM}\sqrt{1-\frac{2GM}{c^2R}} \,.
\end{eqnarray}
These equations govern the eigenmode frequencies of stellar oscillations within the framework of the relativistic Cowling approximation. A more comprehensive derivation of this approximation is provided in Ref.~\cite{Zhao:2022toc}.

\begin{figure}[h]
\includegraphics[width=\linewidth]{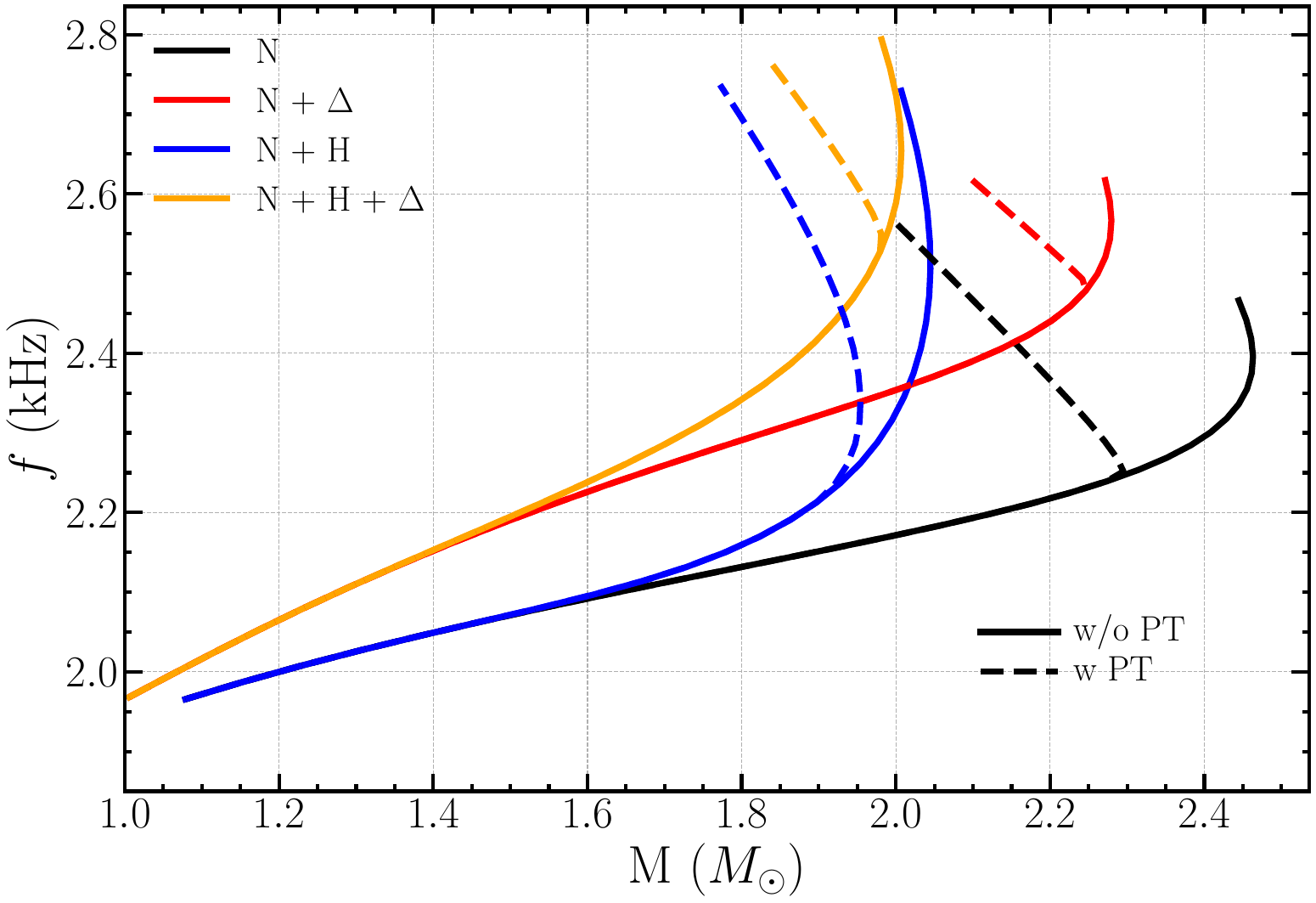}
\caption{Same as Figure~\ref{frequency_mass}, but using Cowling approximation.  }
		\label{frequency_masscow}
        \end{figure}

Figure~\ref{frequency_masscow} shows the $f$-mode frequency–mass relations computed using the Cowling approximation, enabling direct comparison with the full GR results in Figure~\ref{frequency_mass}. For the purely nucleonic EoS without a phase transition, the frequency increases from 2.05 kHz at 1.4\,$M_{\odot}$ to 2.37 kHz at the maximum mass. When a phase transition is included, the frequency at maximum mass decreases to 2.24 kHz. As expected, the Cowling approximation consistently overestimates the $f$-mode frequencies compared to full GR calculation, with the relative error decreasing from 27\% at 1.4\,$M_{\odot}$ to 11.93\% at the maximum mass for the nucleonic case. However, for EoSs with a phase transition, the relative error at maximum mass increases compared to the purely hadronic case. This trend is consistent across all EoS models studied, as summarized in Table~\ref{tab:frequency_comparison}, and confirms that discrepancies between the two methods are more pronounced at lower masses and tend to diminish with increasing mass - except in the presence of a phase transition, where the error becomes more pronounced again near the maximum mass.

\begin{figure}[h]
\includegraphics[width=\linewidth]{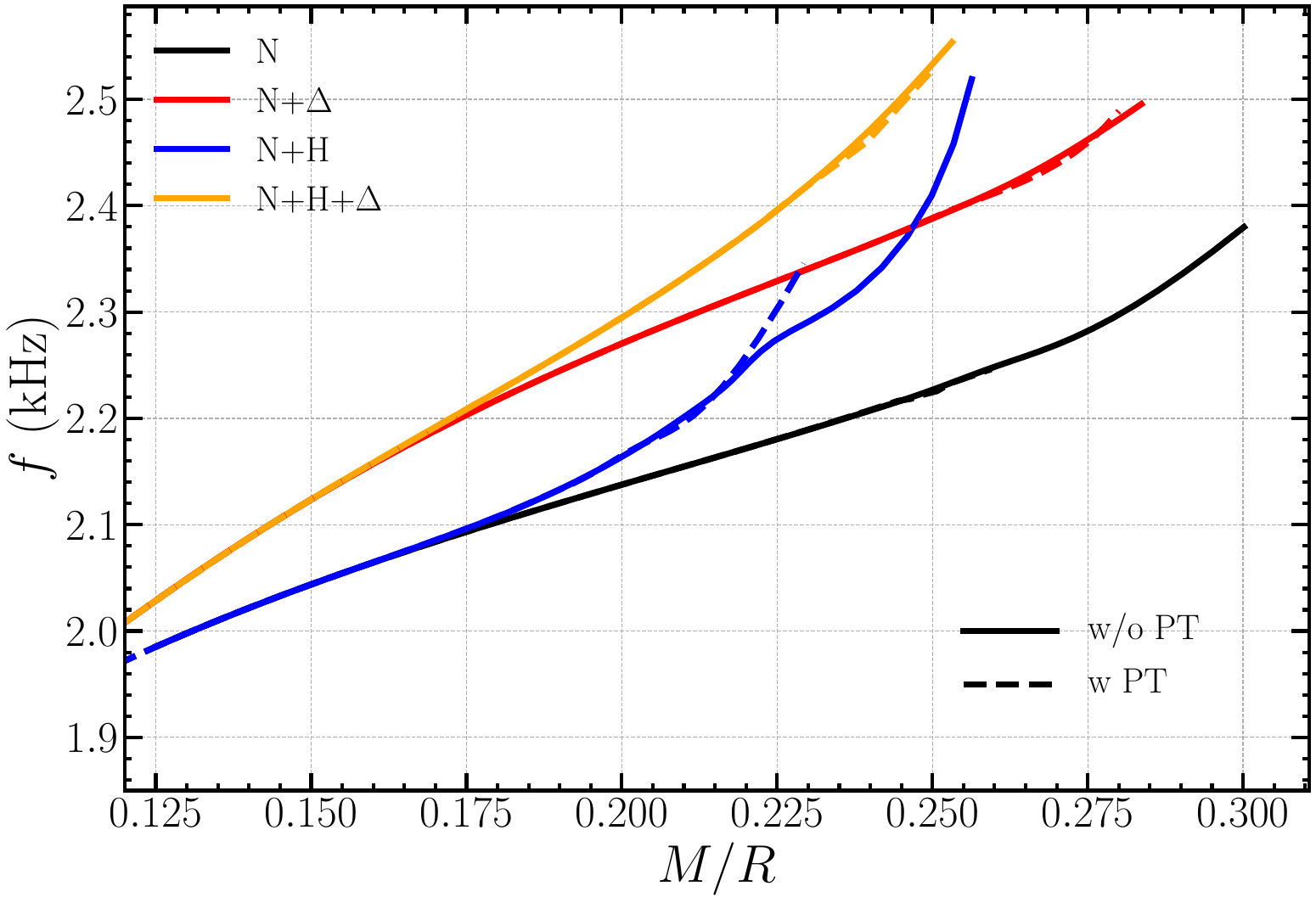}
\caption{Same as Figure~\ref{figCf}, but using Cowling approximation.  }
		\label{Cf_cow}
        \end{figure}

\begin{figure}[h]
\includegraphics[width=\linewidth]{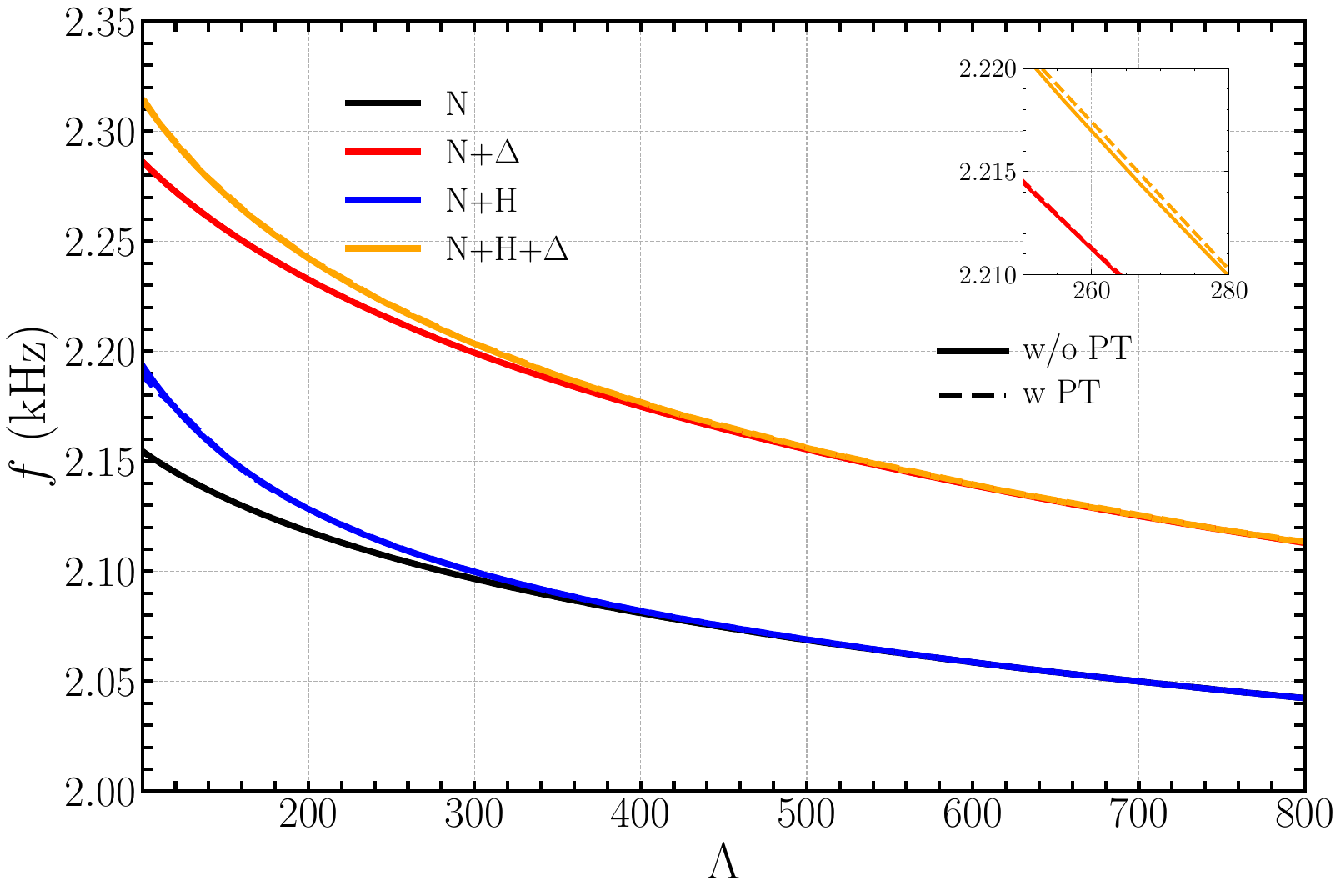}
\caption{Same as Figure~\ref{figLambdaf}, but using Cowling approximation. }
		\label{lamf_cow}
        \end{figure}

Figure~\ref{Cf_cow} presents the $f$-mode frequency as a function of stellar compactness computed under the relativistic Cowling approximation. Compared to the full GR results shown in Figure~\ref{figCf}, all curves exhibit a systematic upward shift in frequency. This offset, typically in the range of 0.3–0.5 kHz, arises due to the omission of gravitational back-reaction, which effectively increases the stiffness of the restoring force in the oscillation equations. Despite this quantitative discrepancy, the Cowling approximation retains the qualitative features of the GR treatment and the impact of the phase transition to quark matter is clearly visible through the dashed segments that deviate from the solid lines near the maximum compactness. Notably, the relative error introduced by the Cowling approximation is larger at low compactness, reaching up to $\sim 30\%$, but decreases to $\sim 10\text{--}15\%$ near the maximum-mass configurations. This trend suggests that while the Cowling approximation may overestimate absolute frequencies, it remains a useful and computationally efficient tool for studying massive neutron stars, especially in exploratory analyses where full GR treatment is computationally expensive.

Figure~\ref{lamf_cow} shows the $f$-mode frequencies as a function of the dimensionless tidal deformability computed using the Cowling approximation. The frequencies remain within the range inferred from the GW170817 event. Similar to the full GR results, we observe a convergence of $f$-mode frequencies beyond $\Lambda \approx 300$, where differences across various EoSs, including those with and without a phase transition to quark matter, become negligible. This suggests that at high deformability, the Cowling approximation also loses sensitivity to compositional effects. The inset highlights the low-$\Lambda$ region, where distinctions between EoSs with and without a phase transition are more apparent.

%

\end{document}